\documentclass{article}

\usepackage{arxiv}

\usepackage[utf8]{inputenc} 
\usepackage[T1]{fontenc}    
\usepackage{hyperref}       
\usepackage{url}            
\usepackage{booktabs}       
\usepackage{amsfonts}       
\usepackage{nicefrac}       
\usepackage{microtype}      
\usepackage{lipsum}		
\usepackage{graphicx}
\usepackage{natbib}
\usepackage{doi}
\usepackage{makecell}

\title{Precision Mental Health: Predicting Heterogeneous Treatment Effects for Depression through Data Integration}

\author{
 Carly Lupton Brantner \\
  Department of Biostatistics\\
  Johns Hopkins Bloomberg School of Public Health\\
  \texttt{clupton1@jhu.edu} \\
   \And
 Trang Quynh Nguyen \\
 Department of Mental Health\\
 Johns Hopkins Bloomberg School of Public Health
  \And
  Harsh Parikh \\
  Department of Biostatistics\\
  Yale University
  \And
  Congwen Zhao \\
  Department of Biostatistics and Bioinformatics\\
  Duke University
  \And
  Hwanhee Hong\\
  Department of Biostatistics and Bioinformatics\\
  Duke University
  \And
  Elizabeth A. Stuart \\
  Department of Biostatistics\\
  Johns Hopkins Bloomberg School of Public Health
 \\
}

\renewcommand{\shorttitle}{\textit{Predicting Treatment Effects in Depression}}

\hypersetup{
pdftitle={A template for the arxiv style},
pdfsubject={q-bio.NC, q-bio.QM},
pdfauthor={David S.~Hippocampus, Elias D.~Striatum},
pdfkeywords={First keyword, Second keyword, More},
}

\begin{document}
\maketitle

\begin{abstract}
When treating depression, clinicians are interested in determining the optimal treatment for a given patient, which is challenging given the amount of treatments available. To advance individualized treatment allocation, integrating data across multiple randomized controlled trials (RCTs) can enhance our understanding of treatment effect heterogeneity by increasing available information. However, extending these inferences to individuals outside of the original RCTs remains crucial for clinical decision-making. We introduce a two-stage meta-analytic method that predicts conditional average treatment effects (CATEs) in target patient populations by leveraging the distribution of CATEs across RCTs. Our approach generates 95\% prediction intervals for CATEs in target settings using first-stage models that can incorporate parametric regression or non-parametric methods such as causal forests or Bayesian additive regression trees (BART). We validate our method through simulation studies and operationalize it to integrate multiple RCTs comparing depression treatments, duloxetine and vortioxetine, to generate prediction intervals for target patient profiles. Our analysis reveals no strong evidence of effect heterogeneity across trials, with the exception of potential age-related variability. Importantly, we show that CATE prediction intervals capture broader uncertainty than study-specific confidence intervals when warranted, reflecting both within-study and between-study variability.
\end{abstract}



\maketitle

\section{Introduction}
When treating patients with depression, clinicians are often interested in ``what works for whom,'' \citep{roth2006works} meaning identifying which treatment works best for a particular individuals given their characteristics. Understanding this aids optimal allocation of resources and increases efficiency, as well as improves outcomes for target patients. The National Institute of Mental Health highlighted the importance of precision medicine -- sometimes also called personalized medicine -- in their 2024 strategic plan, striving for optimally matching treatments with particular patients \citep{NIMH_2024}. The field of precision medicine is of high interest across healthcare, and many have noted the potential far-reaching impact alongside the challenges that come with making decisions at the individual level \citep{dzau2016realizing,kosorok2019precision}.

One specific psychiatric disorder in which precision medicine is of interest is major depressive disorder (MDD), a condition that has a lifetime prevalence ranging from 2 to 21\% worldwide \citep{Sheffler_Patel_Abdijadid_2023}. Antidepressants are a common treatment for MDD; they make up a large and growing drug class in which no drug has been labeled the first choice for all patients \citep{currie2020understanding} with more than 30 possible pharmacotherapeutic options available \citep{d2015vortioxetine}. Physicians and patients therefore have a wide array of possible options to choose from when treating MDD with an antidepressant, and there is interest in learning whether there may be heterogeneity in the effects across medications based on patient characteristics. 

To make personalized treatment decisions, researchers and practitioners often rely on the estimation of heterogeneous treatment effects -- the effects of treatment conditional on observed characteristics of the patients of interest. Estimation of heterogeneous treatment effects is a common goal when assessing treatment efficacy; however, conclusions are often limited by small sample sizes and by the fact that treatment effect heterogeneity can be due to unknown and complex interactions of characteristics \citep{yusuf1991analysis}. For example, when estimating treatment effects from randomized controlled trials (RCTs), a commonly encountered issue is that the trials are designed (and powered) to estimate an average effect and not to estimate quantities relevant for personalized medicine. Choosing and testing a few pre-specified effect moderators is challenging to do with a single trial, and pre-specifying these covariates might cause unknown or complex subgroup differences to be missed. On the other hand, testing all variables as potential effect moderators can be problematic due to multiple testing concerns. Researchers are often interested in the conditional average treatment effect (CATE) \citep{mills2021detecting}, i.e., the average effect conditional on a set of observed characteristics. This estimand considers treatment effect heterogeneity through the inclusion of multiple effect moderators that can interact with one another; a single RCT is therefore rarely powered to estimate this quantity reliably. 

To learn about effect heterogeneity and estimate the CATE, researchers could potentially address the limited power and generalizability of single studies by combining individual, participant-level data (IPD) from multiple sources, e.g., multiple RCTs. A growing body of literature has focused on data integration methods to leverage the benefits of combining data sources and account for the limitations of single sources on their own \citep{brantner2023review, colnet_causal_2021}. There are many methods to combine RCTs, and meta-analysis is a common and standard approach to do so. Notably, meta-analysis is not typically parameterized to estimate heterogeneous treatment effects and instead generally focuses on estimating the average effect across studies. Some work has investigated meta-analysis in a causal inference setting. Specifically, \cite{sobel_causal_2017} defined a causal framework in meta-analysis to estimate study-specific potential outcomes and assess reasons for heterogeneity across trials, and \cite{dahabreh_towards_2020} focused on transporting causal effects from a meta-analysis to a target setting. To introduce alternatives to meta-analysis, \cite{brantner2024comparison} extended and compared several non-parametric methods for combining multiple trials to estimate the CATE. In most parametric and non-parametric approaches for estimating the CATE using multiple trials, a key consideration is accounting for the fact that the data came from different trials, which often yields CATEs that depend not only on characteristics of patients, but also on trial membership. This variability in CATEs across trials can be the result of differences in the distributions of unobserved effect moderators \citep{post2024flexible}, where trial membership can serve as a proxy for the unobserved differences in covariate distributions. Furthermore, differences in the CATE across trials -- potentially due to such differences in unobserved effect moderator distributions -- can be particularly challenging to deal with, then, when aiming to use the CATE to guide decision-making for a set of patients outside the set of trials.

The primary goal of this paper is to guide treatment decisions in a target setting of patients with depression by \textit{predicting} heterogeneous treatment effects. The patients in the target setting do not appear in any of the considered randomized trials; however, they meet the criteria to receive the treatments under study. In the current work, this set of potential treatments is restricted to two, and we assume that individuals in this target setting have not yet received either treatment. Importantly, since the patients in the target setting do not come from any of the trials used to fit the original CATE model, the method must \textit{predict} the CATE in this new setting. This highlights an important distinction between a common goal in causal inference -- estimation -- and another goal that could be very relevant for decision-making -- prediction. We rely on estimation of heterogeneous treatment effects in multiple RCTs but subsequently focus on prediction of the effects in the target setting and patient population of interest.

Our paper extends two-stage meta-analysis to estimate the uncertainty of the CATE for future patients. In this way, the approach can help ensure that treatment guidelines account for effect heterogeneity when needed, and that there is an understanding of when there is not strong evidence for variation in effects. This, ultimately, is often what proponents of personalized medicine are aiming for -- use of existing study results to predict outcomes and guide treatment decisions for individuals outside the original study samples. 
The approach proposed in this paper builds off of meta-analytic prediction intervals \citep{riley_interpretation_2011}, which are commonly used to estimate a range of potential values for an average effect in a new study. Notably, this idea of a new study can also be considered a ``target setting'' or a set of future patients \citep{inthout2016plea}; we use these concepts interchangeably in this paper. We apply this prediction interval approach to the CATE, and we extend the method to the case when non-parametric methods (i.e., causal forest \citep{athey2019generalized} and Bayesian additive regression trees \citep{hill_bayesian_2011}) are used in the first stage of the meta-analysis to estimate the CATE in each of multiple trials.

We utilize this approach to address the question of who in a given health care system would be better suited to one depression medication versus another. We specifically compare two treatments for major depression, duloxetine and vortioxetine, and we integrate data from four randomized controlled trials \citep{mahableshwarkar_randomized_2013,mahableshwarkar_randomized_2015,boulenger_efficacy_2014,baldwin2012randomised} described in detail in \cite{brantner2024comparison}. We estimate the conditional average treatment effect (CATE) in each of these trials, where the primary outcome is the change in a depressive symptoms score from baseline to the last observed follow-up. We then explore how our approach can be used to predict the CATE in our target group of interest -- patient profiles in electronic health records (EHR) from Duke Health Care System. This group represents patients for whom clinicians are interested in understanding treatment effects to aid with decision-making. The methods described in the following sections are used to form prediction intervals for the CATE in this new setting (i.e., the target group).

In the sections to follow, we introduce the depression medications compared in this paper (Section \ref{appintro}), and we discuss notation and the key assumptions required for integrating data and predicting heterogeneous treatment effects in a target setting (Section \ref{notation}). We then explain the extended, two-stage meta-analysis approach for estimating the conditional average treatment effect (CATE) function using data from multiple trials and subsequently forming prediction intervals for the CATE in the target setting, considering both parametric and non-parametric CATE estimation approaches in the first stage of the meta-analysis (Section \ref{methods}). We then investigate performance in simulations based on real data (Section \ref{sims}) and apply the methods to predict the conditional average treatment effects of vortioxetine versus duloxetine for treatment of patient profiles with major depressive disorder (Section \ref{mdd}). Finally, we discuss conclusions, limitations, and future directions in Section \ref{discussion}.

\section{Optimizing Depression Treatment Selection for Individuals}\label{appintro}

With over 30 antidepressants available in practice \citep{d2015vortioxetine}, an ultimate goal is to guide treatment decision-making across all options. However, many of the medications have not yet been compared to one another, and there are some subgroups of antidepressants for which we do not have clinical equipoise; in other words, some medications are not appropriate for comparing for treatment effect heterogeneity because they would be targeted to mostly distinct sets of patients. Two medications that are considered for similar patients and that have been studied together in trials are duloxetine and vortioxetine. 

Duloxetine is a serotonin-norepinephrine reuptake inhibitor (SNRI) that was approved in 2004 and is commonly used to treat MDD, as well as other disorders such as generalized anxiety disorder and fibromyalgia \citep{Dhaliwal_Spurling_Molla_2023}. Several studies have shown its efficacy compared to placebo \citep{girardi2009duloxetine,goldstein2004duloxetine}, and a common dosing of duloxetine for treatment of MDD is 60mg per day, with an option to start the first week at 30mg per day. Vortioxetine is a newer medication, approved in 2013, that has a different mechanism of action; specifically, it involves direct modulation of serotonin receptors \citep{d2015vortioxetine}.

Multiple trials have previously included both duloxetine and vortioxetine as treatment arms \citep{mahableshwarkar_randomized_2013,mahableshwarkar_randomized_2015,baldwin2012randomised,boulenger_efficacy_2014}, but the focus of those trials was mainly on comparing vortioxetine with placebo and using duloxetine as an active reference medication. These studies thus did not investigate effect heterogeneity between vortioxetine and duloxetine, and sample sizes were not powered to do so. Li and colleagues explored these trials in a meta-analysis and found that duloxetine had a greater improvement in symptoms compared to vortioxetine \citep{li2016vortioxetine}. They performed a subgroup analysis by dose of vortioxetine but did not investigate treatment effect heterogeneity in depth. Brantner and colleagues (\citeyear{brantner2024comparison}) developed methods to combine data from multiple trials and applied them to investigate effect heterogeneity of duloxetine versus vortioxetine. This analysis found that there was potential heterogeneity in the treatment effect, whereby older individuals had a larger magnitude of difference in depressive score reduction, but the heterogeneity was minimal at most. Notably, while the analysis by \cite{brantner2024comparison} leveraged multiple trials to assess treatment effect heterogeneity, it did not address what these treatment effects might be in a target setting of future patients who did not participate in the trials. In the following sections, we introduce methods that predict treatment effects in a set of future patients based on inferences from multiple trials, and we subsequently use the methods to provide guidance regarding the choice of vortioxetine or duloxetine for the set of patient profiles.

\section{Setup}\label{notation}

Let $A$ represent treatment assignment, where $A \in \{0, 1\}$ is binary. Let $\boldsymbol{X}$ represent individual-level pre-treatment covariates and $Y$ represent the outcome of interest. Under Rubin's causal framework \citep{rubin_estimating_1974}, we define, for $a \in \{0,1\}$, the potential outcome $Y(a)$ as the potential outcome that would be realized if the individual were assigned (possibly contrary to fact) treatment $a$. Then, the conditional average treatment effect (CATE), $\tau(\boldsymbol{X})$, is defined as:
\begin{equation} \label{unicate}
    \tau(\boldsymbol{X}) = \mathbb{E}(Y(1)|\boldsymbol{X}) - \mathbb{E}(Y(0)|\boldsymbol{X}).
\end{equation} 

In this paper, we consider data from multiple ``settings'', in the form of multiple RCTs as well as a target setting (e.g., patients in a health care system), where the individuals in the target setting are not in any of the trials but are individuals for whom we would like to make informed treatment decisions. We therefore introduce a categorical variable $S$ to indicate the study or setting membership of an individual, where $S\in\{1,\dots,K\}$ if the individual belongs to  one of the RCTs, and $S = K+1$ if the individual is in the target sample. In our scenario, the treatment response depends on what study or setting an individual is in; this results in study/setting-specific CATEs:
\begin{equation} \label{sscate}
    {\tau}_s(\boldsymbol{X})
    =\mathbb{E}(Y(1)|\boldsymbol{X},S=s) - \mathbb{E}(Y(0)|\boldsymbol{X},S=s).
\end{equation}

\subsection{Assumptions} \label{assumptions}

To combine data from multiple trials to estimate the CATE, we employ standard causal inference assumptions, including consistency (Assumption \ref{asx2}), unconfoundedness (Assumption \ref{asx1}), and positivity of treatment assignment (Assumption \ref{asx3}) within each trial. These assumptions are generally satisfied in RCTs and have been described in detail elsewhere \citep{brantner2023review}. In this particular setting where we estimate the CATE using multiple trials and subsequently predict in a target setting, we also require the assumption that every covariate profile in the target setting has positive probability of being found in the trials (Assumption \ref{asx5}). 

\begin{figure}[h!]
    \centering
    \includegraphics[width=10cm]{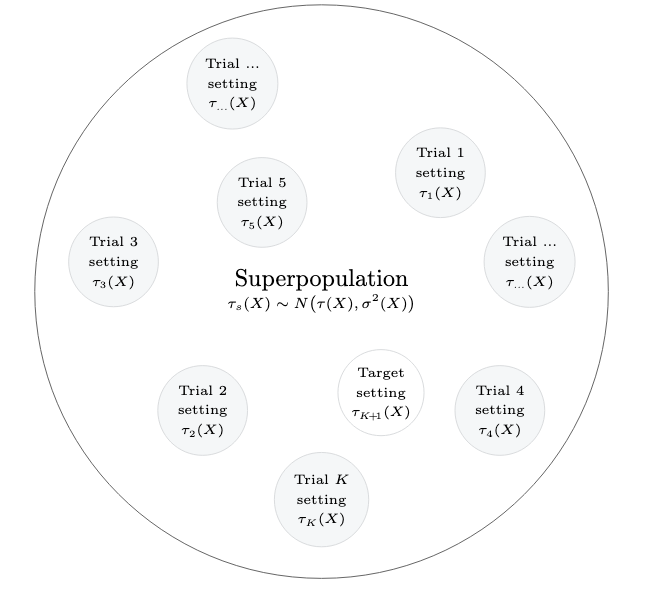}
    \caption{Depiction of overaching assumption of a superpopulation of settings from which we have setting-specific CATEs. We assume that the CATEs for the various settings come from a superpopulation, where the CATEs for a given covariate profile $\boldsymbol{X}$ across settings come from the same distribution. Note that this assumption focuses explicitly on the distribution of $\tau_s(\boldsymbol{X})$, not the distribution of the covariates themselves.}
    \label{fig:assx}
\end{figure}

Furthermore, we assume in this paper that there exists a superpopulation of so-called ``settings'' from which the trials' settings and the target setting are drawn (Figure \ref{fig:assx}). Relatedly, we assume that the CATEs across all trials and the target setting for covariate profile $\boldsymbol{X}^*$ ($\tau_s(\boldsymbol{X}^*)$ as in Equation (\ref{sscate}) for $s \in \{1,\dots,K,K+1\}$) come from the same distribution (Assumption \ref{asxnormal}). The goal in the current work is to predict an interval that contains the CATE given a particular covariate profile in the target setting -- in other words, $\tau_{K+1}(\boldsymbol{X}^*)$.

\newcommand{\indep}{\perp \!\!\! \perp}
\newtheorem{assumption}{Assumption}

\begin{assumption} \label{asx2} 
\textup{[Outcome consistency in trials]}~~$Y = AY(1) + (1-A)Y(0)$ for $S\in\{1,\dots,K\}$.
\end{assumption}
\begin{assumption} \label{asx1} 
\textup{[Within-trial unconfoundedness]}~~$\{Y(0), Y(1)\} \indep A \mid \boldsymbol{X}$ for $S\in\{1,\dots,K\}$.
\end{assumption}
\begin{assumption} \label{asx3}
\textup{[Within-trial positivity]}~~There exists a constant $b>0$ such that~~$b<P(A = 1\mid\boldsymbol{X}=\boldsymbol{x}, S=s)<1-b$~~for each trial $s \in \{1,\dots,K\}$ and for all $\boldsymbol{x}$ values in the trial.
\end{assumption}
\begin{assumption} \label{asx5} 
\textup{[Target setting coverage]}~~$P(S = s\mid\boldsymbol{X}=\boldsymbol{x}^*)>0$~~for all $s \in \{1,...,K\}$ and all $\boldsymbol{x}^*$ present in the target setting.
\end{assumption}
\begin{assumption}\label{asxnormal}
\textup{[Distribution of CATEs across studies/settings]}~~$\tau_s(\boldsymbol{X}^*) \sim \mathcal{N}(\tau(\boldsymbol{X}^*), \theta^2(\boldsymbol{X}^*))$ for $s \in \{1,...,K,K+1\}$.
\end{assumption}




\section{Methods}\label{methods}

In order to predict the CATE for a set of covariate profiles in the target setting, we leverage two-stage meta-analysis extended in two ways: (1) to be applied to the CATE (as opposed to an overall average effect), and (2) to use non-parametric, machine learning approaches in the first stage. We start by describing the standard two-stage meta-analysis, and then introduce our extended usage of the technique to develop CATE prediction intervals.

\subsection{Two-Stage Meta-Analysis}

When combining individual, participant-level data (IPD) across multiple studies, there are two common ways of conducting a meta-analysis: one-stage or two-stage. While one-stage meta-analysis fits one model to all IPD simultaneously (accounting for cross-study heterogeneity when necessary), two-stage meta-analysis first fits study-specific models (Stage 1) and subsequently summarizes aggregate data across those models (Stage 2) 
\citep{riley_two-stage_2021}. Traditionally, both types of meta-analysis are used to summarize information on average treatment effect (ATE), and are shown to perform similarly in previous work \citep{burke_meta-analysis_2017}. 

To move towards CATE estimation and eventual \textit{prediction}, we operationalize a two-stage meta-analysis. We select this version because the two-stage meta-analysis quickly separates within-trial and between-trial heterogeneity \citep{riley_two-stage_2021}, can easily incorporate aggregate-only data, and can be directly extended to use nonparametric methods in the first stage.

A standard two-stage meta-analysis -- typically geared towards estimating the ATE -- starts with Stage 1, in which a model is fit within each trial to produce aggregate data. This model is usually a parametric regression model fit using maximum likelihood estimation \citep{riley_two-stage_2021}, with the model type selected based on the outcome (i.e., linear regression for a continuous outcome, logistic regression for a binary outcome). From this stage, aggregate data is produced across each trial to be summarized in Stage 2. In Stage 2, this aggregate data is combined assuming either a true common treatment effect or random treatment effects. In this paper, we use random effects to allow for between-trial heterogeneity, rather than assuming that there is one common ground truth of the treatment effect across studies. 

\begin{figure}
    \centering
    \includegraphics[width=0.75\linewidth]{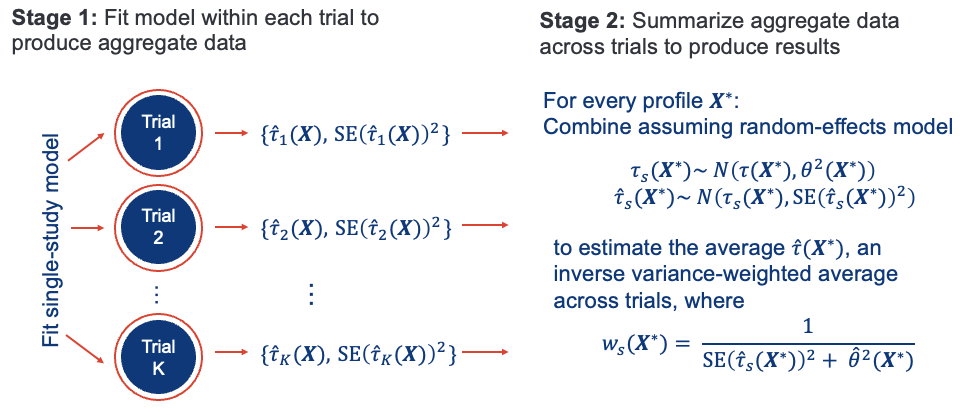}
    \caption{A schematic of two-stage meta-analysis for CATE estimation.}
    \label{fig:twostage}
\end{figure}

We now introduce Stages 1 and 2 for the two-stage meta-analysis as applied to CATE estimation and with nonparametric methods implemented for Stage 1. The processes involved in this approach are illustrated in Figure \ref{fig:twostage}. We then move towards prediction interval generation for this setting. The methods explained here can be implemented using the R package \texttt{multicate}, available on GitHub.

\subsection{Stage 1}

In the first stage of the two-stage meta-analysis for CATE estimation, we estimate the CATE, $\tau_s(\boldsymbol{X})$ for all RCTs $s \in \{1,...,K\}$ separately. This estimation approach is flexible and can be done using a variety of techniques for estimating the CATE in a single study (see \citep{brantner2023review, brantner2024comparison}) -- we introduce three such techniques here.

\subsubsection{Parametric Regression Model}

For a continuous outcome $Y$, CATEs have been traditionally estimated using a parametric regression model with additional interaction terms between treatment and hypothesized effect moderators. We define a parametric linear regression model for a single study during Stage 1 as follows:
\begin{equation}\label{mixefmod}
  \mathbb{E}(Y) = \beta_0^{(s)} + \boldsymbol{\beta}_1^{(s)}\boldsymbol{X} + \beta_2^{(s)}A + \boldsymbol{\beta}_3^{(s)}\boldsymbol{X}^{mod}A
\end{equation}
where $\boldsymbol{X}^{mod}$ represents a (usually low-dimensional) subset of $\boldsymbol{X}$ that consists of hypothesized effect moderators. We use the ${(s)}$ superscript to represent that separate coefficients are estimated according to this model within each study $s \in \{1,...,K\}$. The $\boldsymbol{\beta}^{(s)}$ coefficients represent fixed effects for the model within study $s$, where $\beta_0^{(s)}$ is an intercept term, $\beta_1^{(s)}$ is a vector of fixed effects for covariates $\boldsymbol{X}$, $\beta_2^{(s)}$ is a fixed main effect for treatment $A$, and $\beta_3^{(s)}$ is a vector of fixed interaction effects for treatment-moderator interactions $\boldsymbol{X}^{mod}A$.

In Stage 1, we can fit this regression model once in each of our $K$ studies. We are then interested in the conditional average treatment effect for a covariate profile $\boldsymbol{X}^*$. We can represent the CATE for this covariate profile in study $s$ (as defined broadly in Equation (\ref{sscate})) as \begin{equation}\label{mixefmod_tau}
\begin{aligned}
     {\tau}_s(\boldsymbol{X}^*) &= \mathbb{E}({Y}(1)|\boldsymbol{X}^*, S=s) - \mathbb{E}({Y}(0)|\boldsymbol{X}^*, S=s) \\
    &= {\beta}_2^{(s)} + \boldsymbol{\beta}_3^{(s)}\boldsymbol{X}^{*,mod}.
\end{aligned}
\end{equation} We can therefore estimate $\tau_s(\boldsymbol{X}^*)$ and its variance: $$\hat{\tau}_s(\boldsymbol{X}^*) = \hat{\beta}_2^{(s)} + \boldsymbol{\hat{\beta}}_3^{(s)}\boldsymbol{X}^{*,mod}, \hspace{5mm} SE(\hat{\tau}_s(\boldsymbol{X}^*))^2 = \text{Var}(\hat{\beta}_2^{(s)} + \boldsymbol{\hat{\beta}}_3^{(s)}\boldsymbol{X}^{*,mod}).$$

Notably, this approach requires pre-specification of hypothesized effect moderators, and in this particular case we are also assuming linearity. These requirements of pre-specification are potential drawbacks of estimating the CATE using a parametric approach like linear regression -- this model would not pick up on unknown effect moderators if we did not pre-specify them a priori, and it limits the complexity of the interactions and non-linearities that may be present in the treatment effect heterogeneity. To address these drawbacks, we also explore non-parametric alternatives for estimating the CATE in Stage 1.

\subsubsection{Causal Forest} \label{cf}

As one such non-parametric alternative, the causal forest \citep{athey2019generalized} is a weighted aggregation of causal trees geared specifically towards CATE estimation. Each causal tree is formed by recursively partitioning covariates, where splits are chosen to maximize treatment effect heterogeneity. In each leaf, the treatment effect is estimated as the difference in average outcomes between the treatment and control group individuals within the subgroup that falls in that particular leaf. The causal forest does not rely on estimating the outcomes conditional on covariates and instead directly proceeds to estimation of the treatment effects conditional on covariates.

\cite{wager2018estimation} also introduce a concept called ``honesty'' to their causal forest implementation, which ensures that within each tree, every individual's outcome is used only for defining tree splits or estimating the treatment effect within a leaf, but not both. This concept is discussed in more depth in \cite{wager2018estimation}, and honest versus adaptive (not honest) causal forests are compared briefly in \cite{brantner2024comparison}. This paper provides results from both honest and adaptive causal forests.

In Stage 1, we can apply a single causal forest model to each study. The causal forest is non-parametric so does not provide model coefficients like in meta-analysis. Instead, the model provides CATE estimates and variance estimates per covariate profile and study. Therefore, for a given covariate profile $\boldsymbol{X}^*$ in the target setting, we can calculate estimated means and variances of their treatment effect if they were in each trial: $$\{\hat{\tau}_s(\boldsymbol{X}^*), SE(\hat{\tau}_s(\boldsymbol{X}^*))^2\}.$$

\subsubsection{Bayesian Additive Regression Trees}\label{bart}

Another non-parametric approach for CATE estimation in a single-study is Bayesian Additive Regression Trees (BART) \citep{hill_bayesian_2011, carnegie_examining_2019}. BART is a sum-of-trees model that uses regularization priors to restrict the amount of relationships that each tree can explain. BART is similar to the causal forest in that both are tree-based, but it is different in that it is a Bayesian implementation and focuses on estimating the outcome conditional on covariates rather than the treatment effect. To estimate the CATE using BART, one can estimate the conditional mean outcome under treatment and control, and then directly calculate their difference \citep{hill_bayesian_2011, carnegie_examining_2019}. BART also provides draws from the posterior distributions for outcomes conditional on covariates, so intervals can be created either using the mean and variance of those draws and assuming a normal distribution, or using quantiles of the posterior distribution \citep{dorie_stan_2022}. In this paper, we use a normal distribution assumption to stay consistent with approaches like the causal forest but discuss the alternative of posterior quantiles in the Appendix (\ref{appendix_bart}).

Here, we also fit a single BART model (sometimes called an S-learner \citep{kunzel2019metalearners}) to each study-specific dataset, where covariates include patient-level characteristics and treatment. We estimate each individual's counterfactual according to the fitted BART model by including as ``test'' data the same dataset but with opposite treatment assignment. Finally, we estimate the CATE by subtracting the estimated outcome under treatment minus the estimated outcome under control (averaged across posterior draws), and we estimate the variance of the CATE by adding together the variance of the outcome under treatment across posterior draws with the variance of the outcome under control across posterior draws. Then, just like the causal forest, for a given covariate profile $\boldsymbol{X}^*$ we have: $$\{\hat{\tau}_s(\boldsymbol{X}^*), SE(\hat{\tau}_s(\boldsymbol{X}^*))^2\}.$$ 

\subsection{Stage 2}\label{stage2}

Building off of parametrizations for the ATE \citep{riley_two-stage_2021}, we utilize a random effects model for the second stage of two-stage meta-analysis for CATE estimation for covariate profile $(\boldsymbol{X}^*)$. This random effects model introduces random treatment effects under the assumption that the CATE for covariate profile $(\boldsymbol{X}^*)$ can vary across studies -- in other words, there is not a universal true CATE. This model for Stage 2 can be expressed as follows: 
\begin{equation}\label{assxform}
\begin{aligned}
    &\tau_s(\boldsymbol{X}^*) \sim \mathcal{N}(\tau(\boldsymbol{X}^*), \theta^2(\boldsymbol{X}^*))\\
    &\hat{\tau}_s(\boldsymbol{X}^*) \sim \mathcal{N}(\tau_s(\boldsymbol{X}^*), SE(\hat{\tau}_s(\boldsymbol{X}^*))^2),
\end{aligned}
\end{equation}
where $\tau(\boldsymbol{X}^*)$ represents the average parameter across studies, $\theta^2(\boldsymbol{X}^*)$ represents the between-study variance, and $SE(\hat{\tau}_s(\boldsymbol{X}^*))^2$ represents the within-study variance. Under this assumption, we are interested in estimating the average CATE across studies $\tau(\boldsymbol{X}^*)$, as well as the variance terms, $\theta^2(\boldsymbol{X}^*)$ and $SE(\hat{\tau}_s(\boldsymbol{X}^*))^2$.

We estimate the average CATE for profile $\boldsymbol{X}^*$ across studies, $\tau(\boldsymbol{X}^*)$, using an inverse-variance weighted average: $$\hat{\tau}(\boldsymbol{X}^*) = \frac{\sum_{s=1}^K\hat{\tau}_s(\boldsymbol{X}^*)w_s(\boldsymbol{X}^*)}{\sum_{s=1}^K w_s(\boldsymbol{X}^*)} \hspace{3mm} \text{and} \hspace{3mm} \text{Var}(\hat{\tau}(\boldsymbol{X}^*)) = \frac{1}{\sum_{s=1}^K w_s(\boldsymbol{X}^*)}$$ \text{where} $$w_s(\boldsymbol{X}^*) = \frac{1}{SE(\hat{\tau}_s(\boldsymbol{X}^*))^2 + \hat{\theta}^2(\boldsymbol{X}^*)}.$$

The within-study variance, $SE(\hat{\tau}_s(\boldsymbol{X}^*))^2$ is estimated by the Stage 1 approach, while the between-study variance, $\hat{\theta}^2(\boldsymbol{X}^*)$ can be estimated in a variety of ways, including DerSimonian and Laird's methods of moments estimator \citep{dersimonian1986meta}, other similar techniques, restricted maximum likelihood estimation (REML), and Bayesian inference techniques. In this extension, we utilize REML as recommended in previous work \citep{riley_two-stage_2021}. The weights, $w_s(\boldsymbol{X}^*)$, incorporate both the within-study and between-study variability to balance the precision of the estimates and the heterogeneity of the estimates across studies. Specifically, studies with small within-study variance are up-weighted in the estimation of $\hat{\tau}(\boldsymbol{X}^*)$, but in the presence of high heterogeneity of the CATE estimates across studies, the inclusion of $\hat{\theta}^2(\boldsymbol{X}^*)$ reduces the dominance of very large or precise studies.

\subsection{Prediction Intervals}

From Stages 1 and 2, we have estimated the CATE per trial for a set of covariate profiles $\boldsymbol{X}^*$, and we have summarized such CATEs across trials. Now, we are interested in the treatment effect in the target setting, $\tau_{K+1}(\boldsymbol{X}^*)$. Prediction intervals in random effects meta-analysis estimate a range of potential parameter values in a new setting \citep{riley_interpretation_2011}, so they can be implemented here to determine what the effects might be in the target setting. The general form for a prediction interval is based off of the above assumptions and models and can be expressed as follows: \begin{equation}\label{pi}
    \tau_{K+1}(\boldsymbol{X}^*) \in \Bigl\{\hat{\tau}(\boldsymbol{X}^*) \pm t_{K-2}\sqrt{SE(\hat{\tau}(\boldsymbol{X}^*))^2 + \hat{\theta}^2(\boldsymbol{X}^*)}\Bigl\},
\end{equation} where we can plug in the estimates of $\hat{\tau}(\boldsymbol{X}^*)$, $SE(\hat{\tau}(\boldsymbol{X}^*))^2$, and $\hat{\theta}^2(\boldsymbol{X}^*)$ defined above and use the t-distribution at the $100(1-\alpha/2)$ level with $K-2$ degrees of freedom. The t-distribution is used here to address the uncertainty involved in estimating the between-study variability $\hat{\theta}^2(\boldsymbol{X}^*)$, and $K-2$ degrees of freedom are recommended because both the average CATE for profile $\boldsymbol{X}^*$ across studies, $\hat{\tau}(\boldsymbol{X}^*)$, and the between-study variance, $\hat{\theta}^2(\boldsymbol{X}^*)$, are estimated from the data \citep{higgins_re-evaluation_2009}.

\section{Simulations}\label{sims}

\subsection{Setup}

Before working with the real randomized controlled trials and electronic health record data to compare efficacy of vortioxetine and duloxetine, we present a simulation study to assess performance of the methods described above for forming CATE prediction intervals. We set up the simulation to closely represent the real RCTs discussed in the following Section \ref{mdd} \citep{mahableshwarkar_randomized_2013,mahableshwarkar_randomized_2015,boulenger_efficacy_2014,baldwin2012randomised}; specifically, we used the estimated means and covariances of variables in the real data to guide covariate distributions in the simulated data, and we used treatment effect functions estimated from the real data in the simulated data.

In primary simulations, we simulated $K=10$ studies, each with $n=500$ individuals. Each individual had probability 0.5 of receiving the treatment, and individuals had five observed covariates: sex (defined as binary, female or male), smoking status (defined as binary, have smoked or never smoked), weight, age, and baseline Montgomery-Asberg Depression Rating Scale (MADRS) score \citep{montgomery1979new}. These covariates were simulated using a multivariate normal distribution within each study, with continuous variables simulated on a standardized scale. The covariance matrix for the simulated covariate distributions was estimated using the real data and held constant across iterations. We simulated covariate means according to the following, where $\overline{\text{Covariate}}_s$ represents the mean of the covariate value in study $s$: $\overline{\text{Age}}_s \sim \mathcal{N}(0, 0.2^2); \overline{\text{Sex}}_s \sim \mathcal{N}(0.6784, 0.1^2); \overline{\text{Smoking Status}}_s \sim \mathcal{N}(0.3043, 0.1^2); \overline{\text{Weight}}_s \sim \mathcal{N}(0, 0.5^2); \overline{\text{Baseline MADRS}}_s \sim \mathcal{N}(0, 0.3^2)$. For sex and smoking status, the simulated values were converted to binary. These standard deviations of covariate means across studies were approximated to mimic the standard deviations of covariate means across studies in the real RCTs analyzed in the following section.


While data from the trials were randomly sampled according to the above information for every iteration of the simulation, we also created a \textit{single} target sample with $n=100$ individuals representing a range of covariate profiles. Baseline covariates for the target sample were sampled from a multivariate normal distribution that differed from the distribution the trials were sampled from. Specifically, the target sample distribution was older, had a higher proportion female, had a higher proportion smoking, had higher weight, and had lower baseline depression according to the MADRS. We saved this set of covariate profiles to be used across all simulation iterations and settings. 


Average outcomes for each covariate profile in the training trials and target setting were defined according to the following model: $$Y = m(\boldsymbol{X}) + A*\tau(\boldsymbol{X}) + \epsilon,$$ where $Y$ represents the change in MADRS score from baseline to last observed follow-up, $m(\boldsymbol{X})$ represents a main effect function, $A$ represents treatment (0 representing control and 1 treatment), $\tau(\boldsymbol{X})$ represents the CATE, and $\epsilon \sim \mathcal{N}(0, 0.05^2)$ is a random error term. We considered two settings for $m$ and $\tau$:

\begin{enumerate}
    \item Age is the only moderator, CATE is linear: $$m(\boldsymbol{X}) = (-17.40 + a_s) - 0.13*\text{Age} - 2.05*\text{MADRS} - 0.11*\text{Sex}$$ $$\tau(\boldsymbol{X}) = (2.505+b_s) + (0.82+c_s)*\text{Age}$$
     \item Age is the only moderator, CATE is non-linear: $$m(\boldsymbol{X}) = (-17.52+a_s) -0.08*\text{Age}$$ $$\tau(\boldsymbol{X}) = (2.20 + b_s)*\exp[(0.35 + c_s)*\text{Age}]$$
\end{enumerate}

In these setups, $a_s \sim \mathcal{N}(0, \sigma_a^2)$, $b_s \sim \mathcal{N}(0, \sigma_b^2)$, and $c_s \sim \mathcal{N}(0, \sigma_c^2)$ represent heterogeneity due to study membership. We included different values for the standard deviations of these study-level terms to allow for varying heterogeneity in the main and treatment effects across studies. Specifically, we used three sets of values:

\begin{enumerate}
    \item Heterogeneous intercept: $\sigma_a = 1$, $\sigma_b = 0.25$, $\sigma_c = 0.25$
    \item Heterogeneous intercept and main treatment effect: $\sigma_a = 1$, $\sigma_b = 0.5$, $\sigma_c = 0.25$
    \item Heterogeneous intercept, main treatment effect, and treatment-moderator interaction: $\sigma_a = 1$, $\sigma_b = 1$, $\sigma_c = 0.5$
\end{enumerate}

While we generated new values for $a_s$, $b_s$, and $c_s$ across trials for every iteration, we only generated $a_{\text{target}}$, $b_{\text{target}}$, and $c_{\text{target}}$ once for each of the three levels of heterogeneity in the target sample. 


Finally, we reran the above settings with a few other changes to the parameters. First, we allowed for two different settings in terms of the covariate distributions; one setting had the same mean covariate value across trials as opposed to variable means, and another setting only had variable mean ages across trials. Secondly, we reran the above simulations with $K=4$ and $K=30$ RCTs. Finally, we reran the above simulations where instead of $a_s \sim \mathcal{N}(0, \sigma_a^2)$, $b_s \sim \mathcal{N}(0, \sigma_b^2)$, and $c_s \sim \mathcal{N}(0, \sigma_c^2)$, we introduced non-normal (uniform) distributions across trials: $a_s \sim U(-1, 1)$, $b_s \sim U(-1, 1)$, and $c_s \sim U(-1, 1)$.

For each simulation setup, we ran 500 replications. The performance of the methods was assessed based on prediction interval coverage, prediction interval length, and bias. Coverage for each patient profile in the target setting was calculated as the percentage of the iterations for which the covariate profile's true treatment effect was contained within their prediction interval. Prediction interval length was calculated as average the average length of the prediction interval across iterations for each covariate profile separately, and ``bias'' was calculated as the average difference between the center of the prediction interval and the true CATE in the target setting across iteractions for each covariate profile separately.

We used several R packages to conduct the simulations, including \texttt{metafor} for the two-stage meta-analysis \citep{metafor}, \texttt{grf} for the causal forest \citep{athey2019generalized}, and \texttt{dbarts} for BART \citep{dorie2023package}. In the causal forest and BART, hyperparameters were set to be the defaults, except that the causal forest was set to use 1,000 trees instead of the default of 2,000 for computational ease. Code containing all methods and implementation of the simulations can be found at the repository: \verb|https://github.com/carlyls/Predict_CATE| and the estimation and prediction methods are incorporated in the R package \verb|multicate| available on GitHub.

\subsection{Results}

We present results from the primary simulations, including $K=10$ RCTs with $n=500$ individuals in each and variable covariate distributions across trials. Results from both a linear and non-linear CATE with varying levels of heterogeneity in the coefficients across trials are included. 

\begin{figure}[h!]
    \centering
    \includegraphics[width=17cm]{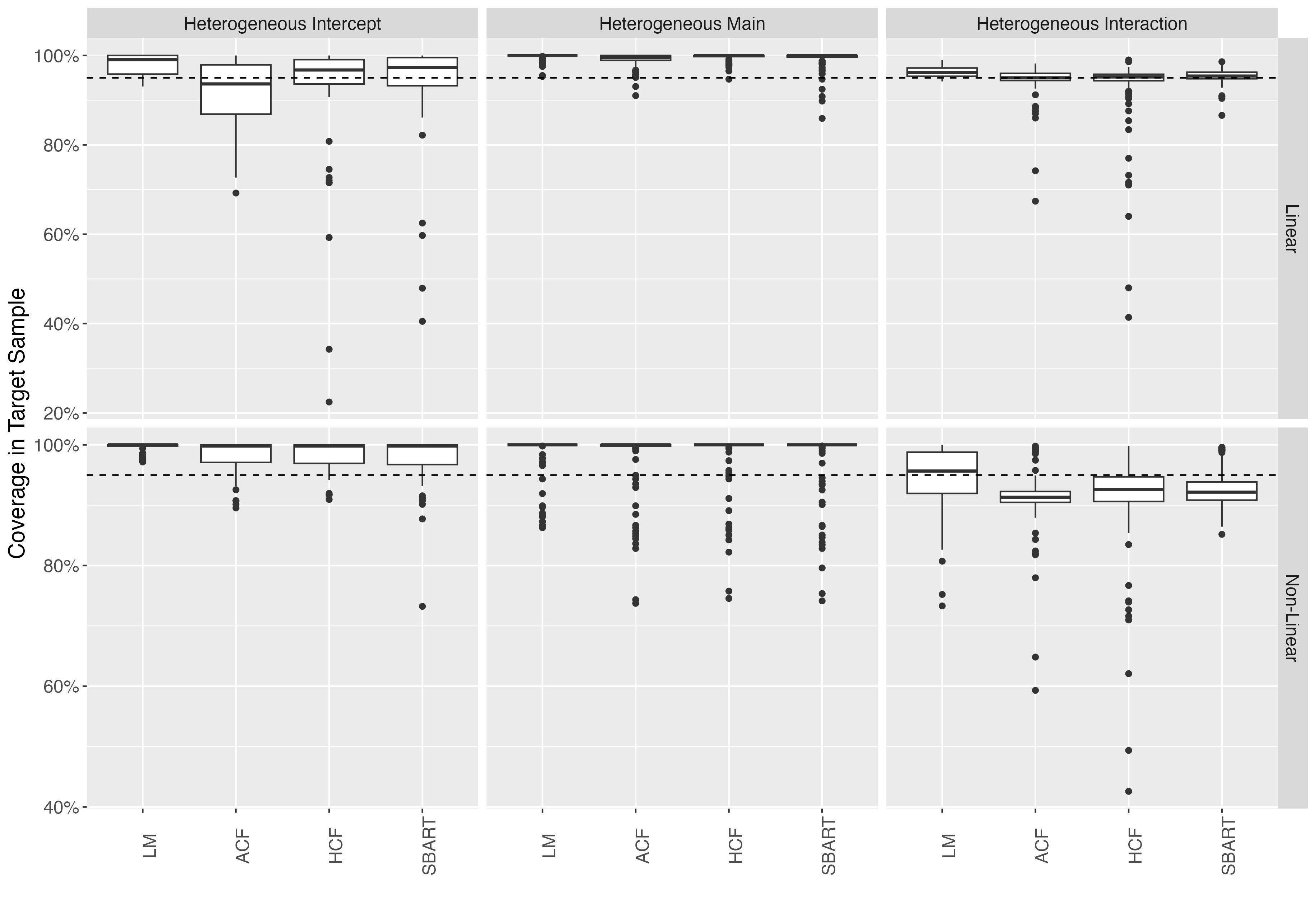}
    \caption{Distributions of coverage for each covariate profile in the target setting across each method and data generation scenario, where coverage was calculated as the percent of 500 iterations for which the profile's true treatment effect was contained within the estimated prediction interval.\\
    \textit{Method abbreviations:} MA = meta-analysis, ACF = adaptive causal forest, HCF = honest causal forest, SBART = Bayesian Additive Regression Trees with S-learner.}
    \label{fig:coverage}
\end{figure}

Figure \ref{fig:coverage} displays boxplots of CATE prediction interval coverage for the 100 covariate profiles in the target setting. Overall, there were high levels of coverage for the target setting profiles across the majority of methods and simulation scenarios; specifically, most profiles averaged around 95\% or higher coverage. Coverage seemed to be very similar across the four Stage 1 methods explored (linear model, adaptive causal forest, honest causal forest, and BART with an S-Learner). While the coverage for some settings centered at around 95\% (linear CATE with heterogeneous intercept, linear CATE with heterogeneous interaction, and non-linear CATE with heterogeneous interaction), the coverage for others was closer to 100\% (non-linear CATE with heterogeneous intercept, linear CATE with heterogeneous main effect, and non-linear CATE with heterogeneous main effect). We explore these results more by investigating interval length.

\begin{figure}[h!]
    \centering
    \includegraphics[width=17cm]{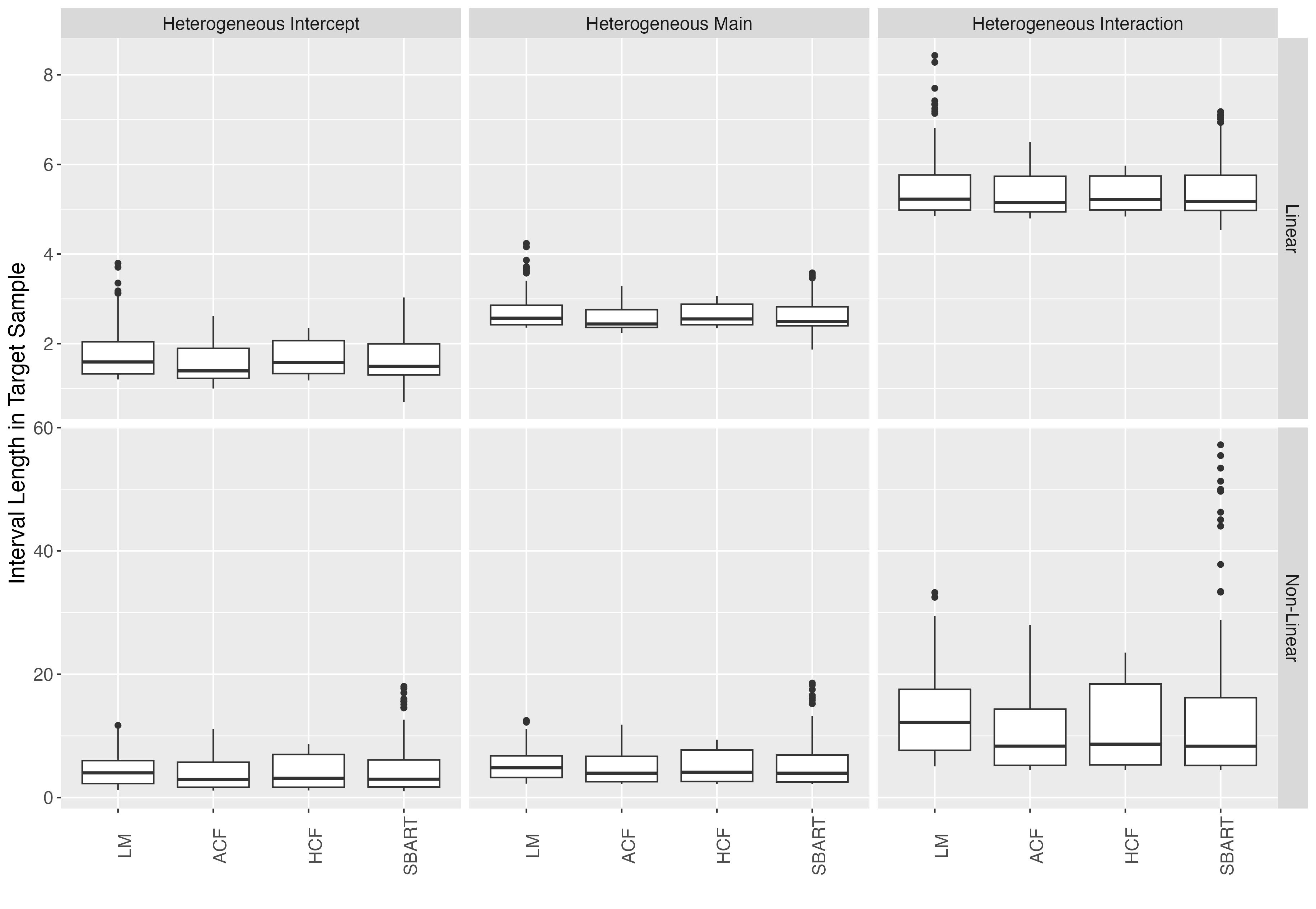}
    \caption{Distributions of average interval length for each covariate profile in the target setting across each method and data generation scenario, where length was calculated as the average length of the profile's prediction interval across 500 iterations.\\
    \textit{Method abbreviations:} MA = meta-analysis, ACF = adaptive causal forest, HCF = honest causal forest, SBART = Bayesian Additive Regression Trees with S-learner.}
    \label{fig:length}
\end{figure}

Figure \ref{fig:length} presents the average interval length for each covariate profile in the target setting. Again here, all methods yield similar results. For all methods, higher length occurred when heterogeneity across trials increased. The results for bias are included in the Supplementary Material and broadly display that bias was similar across methods. Notably, the center of the prediction interval for the CATE for covariate profile $\boldsymbol{X}^*$ is an estimator of the mean CATE for profile $\boldsymbol{X}^*$ across studies – it is not explicitly meant to be an estimator of the true CATE for profile $\boldsymbol{X}^*$ in the target setting. Instead, we rely on the prediction interval to cover the true CATE for profile $\boldsymbol{X}^*$ in the target setting. We therefore emphasize that coverage is the goal of the approach in this paper, as opposed to ``bias'' as defined as the difference between the center of the prediction interval and the true CATE for $\boldsymbol{X}^*$ in the target setting.

To further examine the variability in metrics like coverage, interval length, and absolute bias, we plotted these metrics versus values of age across the covariate profiles in the target setting (see Supplement). The goal here was to determine whether the methods performed equally well across all values of the covariate or not. In terms of coverage, all methods had consistently high coverage for profiles with age close to the mean. For some settings, all methods had lower coverage for profiles with more extreme values of age (such as three standard deviations from the mean).

Interval length also varied by age and data generation setup; specifically, interval length was higher for profiles with age further from the mean. This relationship was most notable for linear models in the linear CATE setting, and BART in the non-linear CATE setting. In the high heterogeneity, non-linear CATE setting, some profiles had very wide prediction intervals.

In the Supplement, we also display coverage results for a few alternative simulation settings. When we switched from variable covariate distributions across studies to the same covariate distributions or just variable age across studies, the coverage results remained largely consistent with the primary simulation results. Coverage was also consistent across number of studies, $K$, of 4, 10, and 30, as shown in the Supplement. Finally, coverage remained mostly high but below 95\% when the CATE coefficients followed a uniform, as opposed to normal, distribution across studies.

\section{Predicting the CATE for Major Depression Treatments} \label{mdd}

\subsection{Datasets}

We now compare efficacy of two treatments for major depression: duloxetine and vortioxetine. We applied the approach introduced above, utilizing data from four trials comparing these two treatments to then predict in a target setting represented by electronic health record data. 

Each trial included participants who were between 18 to 75 years old, had a Major Depressive Episode (MDE) as a primary diagnosis according to the DSM-IV-TR criteria over at least three months, and had a Montgomery-Asberg Depression Rating Scale (MADRS) \citep{montgomery1979new} score of at least 22 (one trial) or 26 (three trials) at both screening and baseline \citep{mahableshwarkar_randomized_2013,mahableshwarkar_randomized_2015,boulenger_efficacy_2014,baldwin2012randomised}. Participants were randomly assigned to receive duloxetine, vortioxetine, or placebo; we removed individuals who were randomly assigned to placebo for this analysis. We treated duloxetine as the reference condition here because it was already in use at the time of the trials. The primary outcome of interest was the change in MADRS score from baseline to last observed follow-up, where the goal was to follow patients for 8 weeks. More information on these trials can be found in their original papers \citep{mahableshwarkar_randomized_2013,mahableshwarkar_randomized_2015,boulenger_efficacy_2014,baldwin2012randomised} or in \cite{brantner2024comparison}. Table~\ref{tab:desc} presents descriptive statistics for the four trials used in the current analysis.

We constructed our external target setting using data from patients at the Duke Health Care System. We identified patients from Duke psychiatry or primary care with visits for major depression, bipolar disorder, and/or persistent mood disorder, and we filtered to patients who were prescribed either duloxetine or vortioxetine between January 1, 2014 to December 31, 2021. We then subset to patients who had at least one year of EHR data before they were prescribed either of the medications, and we finally included only patients aged 18 to 65 at the time of their prescription and patients who had a non-missing PHQ-9 score of at least 10 (indicating at least moderate depression) \citep{kroenke2001phq}. The final sample size from the EHR data was 2,172 patients. We used these patients to represent a set of possible covariate profiles in the target setting, with a goal to predict the CATE for this set of profiles. For the purposes of this analysis, we ignored which treatment the patients actually received and used their baseline characteristics to form treatment effect prediction intervals based on the approaches previously described. Descriptive statistics for this patient sample can also be found in Table~\ref{tab:desc}. Notably, the EHR sample was similar to the trials in terms of age, weight, sex, and baseline depression; however, the EHR data had much higher prevalence of the comorbidities and medications measured.

\bgroup
\def\arraystretch{1.5}
\begin{table}[h]
    \centering
    \begin{tabular}{|rrrrrr|}
         \hline
         & \textbf{NCT00635219} & \textbf{NCT00672620} & \textbf{NCT01140906} & \textbf{NCT01153009} & \textbf{Duke Patients} \\
         & N=575 & N=418 & N=436 & N=418 & N=2,172\\
         
         & \textit{Mean (SD)} & \textit{Mean (SD)} & \textit{Mean (SD)} & \textit{Mean (SD)} & \textit{Mean (SD)} \\
         
         Age & 46.3 (11.5) & 43.0 (13.8) & 46.3 (13.9) & 43.4 (12.2) & 44.8 (12.7) \\
         Weight (lbs) & 156 (35.2) & 193 (53.3) & 163 (34.7) & 193 (51.4) & 201 (58.5) \\
         
         & \textit{\%} & \textit{\%} & \textit{\%} & \textit{\%} & \textit{\%} \\
         
         Female & 67.7 & 64.4 & 65.4 & 74.2 & 75.4 \\
         Diabetes Mellitus & 2.1 & 4.8 & 1.4 & 2.4 & 20.7\\
         Hypothyroidism & 2.8 & 5.0 & 3.2 & 4.5 & 11.1\\
         Anxiety & 3.7 & 1.9 & 0.2 & 3.8 & 61.0\\
         Antidepressant & 24.3 & 1.7 & 33.5 & 19.4 & 58.2\\
         Thyroid Medication & 2.6 & 0.7 & 3.4 & 3.6 & 9.3\\
         \makecell{Severe (vs. Moderate) \\Baseline Depression} & 24.9 & 13.2 & 17.2 & 28.7 & 26.8 \\
         \hline
    \end{tabular}
    \vspace{2mm}
    \caption{Descriptive statistics for four randomized controlled trials and EHR data from patients at the Duke Health Care System.}
    \label{tab:desc}
\end{table}
\egroup

Across both samples, conditional mean imputation was performed for missing values of weight (n=1 in the trials and n=16 in the EHR data). Most variables were similarly defined across the trials and the EHR data; however, the RCTs used MADRS to measure depression severity, while the EHR data used PHQ-9. In order to have a similar measure of baseline depression across all datasets, we created a binary indicator of moderate or severe depression based on criteria defined for both scales \citep{snaith1986grade, herrmann1998sunnybrook, kroenke2001phq}.

\subsection{Results}

After pre-processing the data, we applied the two-stage meta-analytic approach described previously to first fit CATE models to the four RCTs (Stage 1) and subsequently use a meta-analysis (Stage 2) to form treatment effect prediction intervals for the covariate profiles in the EHR data. Here, we focus on the results from the honest causal forest used in Stage 1. 


The average CATE estimate across the 1,847 individuals in the trials was 2.32 (SD = 1.45). The average confidence interval length in the trials was 5.74, and 33.62\% of the sample had a confidence interval that did not cross zero. The majority of covariate profiles (n=1,774, 96.05\%) represented in this sample according to this model had a positive treatment effect estimate, thus in favor of duloxetine. There was potential heterogeneity of the treatment effect by age, where older individuals had higher estimates of the effect, but this heterogeneity seemed minimal.

\begin{figure}[h!]
    \centering
    \includegraphics[width=14cm]{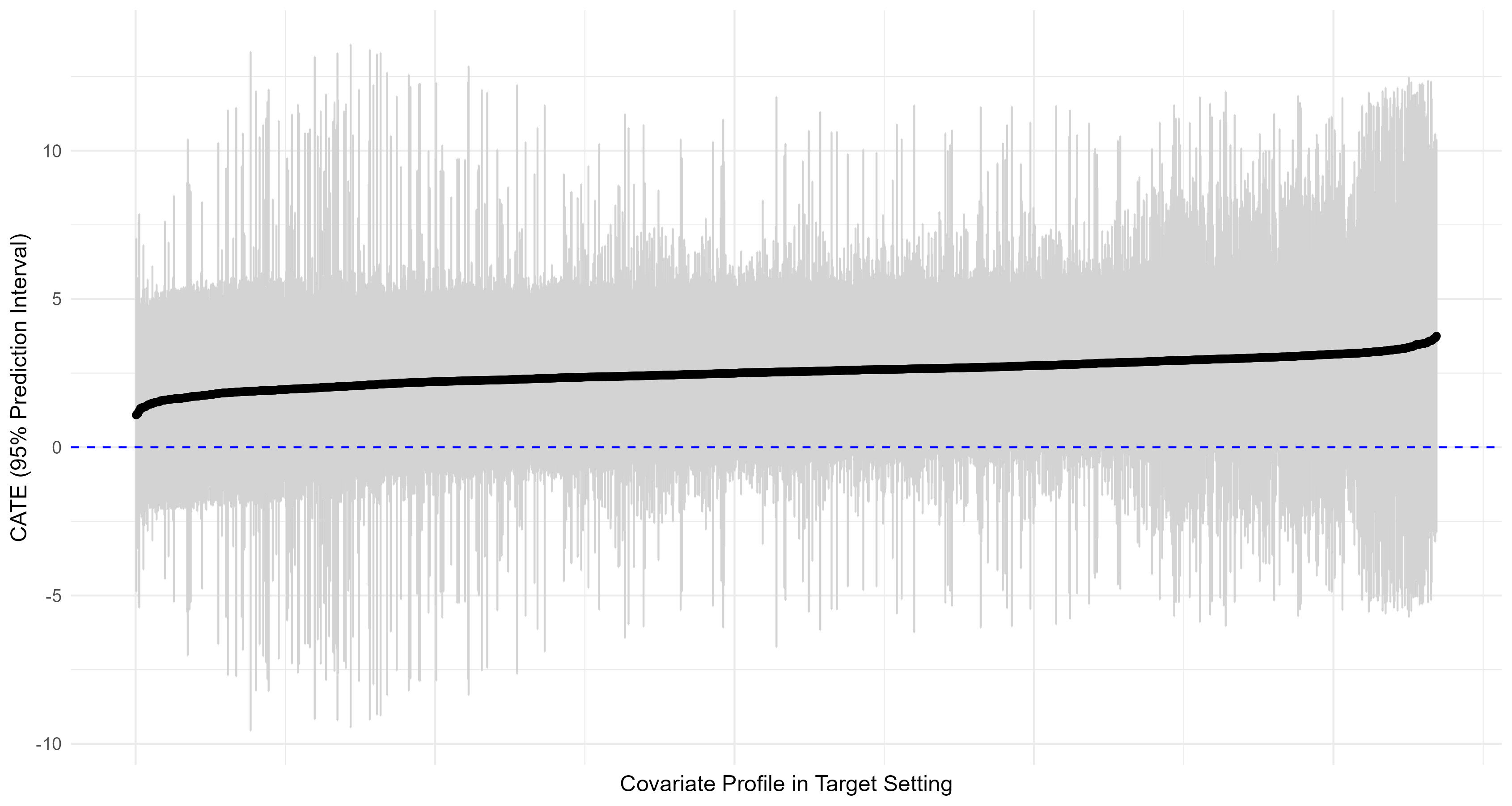}
    \caption{95\% prediction intervals for treatment effects in target setting.*\\
    *X axis is ordered by magnitude of treatment effect. CATE estimates represent the difference in the change in MADRS score from baseline to last observed follow up between vortioxetine and duloxetine. A positive treatment effect therefore indicates that duloxetine was estimated to be more effective than vortioxetine for that particular covariate profile.}
    \label{fig:pis}
\end{figure}

Figure \ref{fig:pis} displays the effect estimates and 95\% prediction intervals for the covariate profiles in the target setting of patients from the EHR data. All EHR patient profiles had positive effect predictions, again indicating that duloxetine was predicted to be the better treatment for reduction of depressive symptoms in this patient population. Specifically, the average CATE estimate in the target setting was 2.52 (SD = 0.45). 1.75\% of the covariate profiles in this target setting had a prediction interval that did not cross zero, and the average interval length was 8.51. 


To investigate the prediction intervals more, Table \ref{tab:sig} displays the EHR patient profiles broken down by whether the associated prediction interval crossed zero or not. Notably, no prediction intervals were significant and in favor of vortioxetine; instead, all either crossed zero or were entirely positive, which indicates that duloxetine was estimated to be more effective. We see that only 38 individuals had a significant prediction interval; a major reason for this is likely due to the t-value used in the calculation of the prediction intervals. Specifically, since there were four studies being combined, the t-value used was from a distribution with $4-2=2$ degrees of freedom, allowing for high levels of uncertainty. From this table, we see that the group for whom the prediction interval did not cross zero had a higher average age, lower average weight, lower percentage with severe depression, and higher percentage of comorbidities (diabetes mellitus, hypothyroidism, and anxiety). Notably, these results are primarily descriptive, and we are not explicitly testing differences across the subgroups displayed in this table.

\bgroup
\def\arraystretch{1.5}
\begin{table}[h!]
    \centering
    \begin{tabular}{|rrr|}
    \hline
    & \textbf{Prediction Interval Crossed 0} & \textbf{Duloxetine Significantly Better} \\
    & N = 2,134 & N = 38 \\
    
    & \textit{Mean (SD)} & \textit{Mean (SD)} \\

    Age & 44.7 (12.8) & 48.1 (4.2) \\
    Weight (lbs) & 201.5 (59.0) & 186.8 (19.1) \\

    & \textit{\%} & \textit{\%} \\

    Female & 75.4 & 78.9 \\
    Diabetes Mellitus & 20.6 & 23.7\\
    Hypothyroidism & 11.0 & 15.8 \\
    Anxiety & 60.9 & 68.4 \\
    Antidepressant & 58.1 & 65.8 \\
    Thyroid Medication & 9.3 & 7.9 \\
    \makecell{Severe (vs. Moderate) \\Baseline Depression} & 27.2 & 7.9\\
    \hline
    \end{tabular}
    \vspace{2mm}
    \caption{Descriptive statistics for EHR patient profiles broken down by whether the prediction interval for the CATE crossed 0 or was all greater than 0.}
    \label{tab:sig}
\end{table}
\egroup

\begin{figure}
    \centering
    \includegraphics[width=14cm]{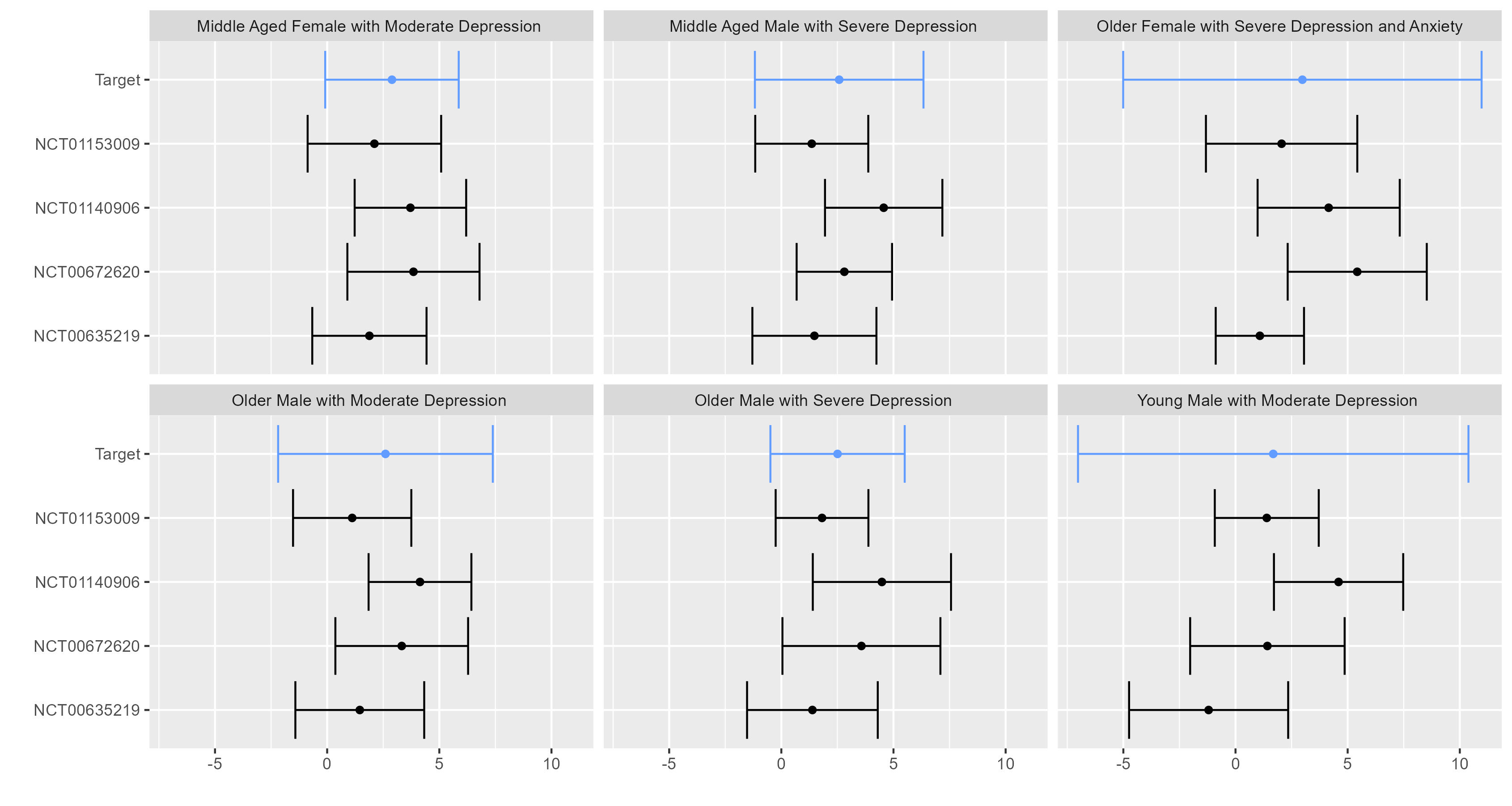}
    \caption{Study-specific confidence intervals and target prediction interval for a random subset of six covariate profiles in the Duke Health Care System.}
    \label{fig:interval_comp}
\end{figure}

Finally, Figure \ref{fig:interval_comp} displays differences between confidence intervals created within individual studies and the prediction interval for the target setting that leveraged information from all available studies. Here, we display a random sample of six covariate profiles and label them by some of their defining profile features, notably not including all characteristics such as comorbidities and medications that may be affecting CATE estimates and prediction interval widths. Overall, we can see in this figure that the prediction interval effectively covers most of the individual study confidence intervals and allows for a higher level of uncertainty due to study-level heterogeneity. This figure depicts the risk of false precision if confidence intervals from a single study are used to draw conclusions, but also the opportunity of increased precision when there is less heterogeneity in the CATE across studies. This exercise exemplifies a somewhat simplified version of what could support decision-making in practice, where we investigate the CATE and its uncertainty for a particular profile across various studies, and then extend the results by forming prediction intervals for the target setting.

From this analysis, we were able to obtain prediction intervals for the conditional average treatment effect in a representative set of patient profiles with depression in a health care system. These prediction intervals can be leveraged to make treatment decisions, and they can also be used to interpret the level of uncertainty around those decisions in the real world setting.

\section{Discussion} \label{discussion}

To advance the field of precision mental health, this paper begins to bridge the gap between information that can be learned from randomized trials to a real-world target patient population. We utilize models built using multiple RCTs and two-stage meta-analysis to develop prediction intervals for the conditional average treatment effect in patients with major depression in a healthcare system. We discuss this approach in conjunction with parametric linear models for the first stage of meta-analysis as well as with non-parametric methods, including the causal forest and BART. 

In this paper, we performed a comparison of vortioxetine and duloxetine to investigate treatment effect heterogeneity. We combined four RCTs to estimate the CATE and subsequently predict CATEs for health records of patients in a health care system. Our analysis revealed that there was potential heterogeneity of the treatment effect by age but that variability was high, and in general, all patient profiles were estimated to benefit more from duloxetine compared to vortioxetine. This analysis could be helpful for clinicians looking to decide between treatment options for patients based on previous treatment effect heterogeneity estimated from randomized trials. These results again demonstrated that duloxetine would be preferable overall; however, other treatment comparisons might reveal more notable differences in treatment effectiveness across patient characteristics. In general, this approach allowed prediction of effects in the target patient sample without needing to observe any outcomes or treatment assignment in this group beforehand. Furthermore, the prediction intervals alleviate the risk of false precision that we might obtain if we relied on simply the confidence intervals from a single trial to make conclusions (Figure \ref{fig:interval_comp}).

It is important to note that in two of the original trials (NCT01140906 and NCT01153009), duloxetine was included as a reference medication, and patients were not included in the study if they had previously not responded to duloxetine. Vortioxetine was investigational at the time of the trials and so the same exclusion criteria did not apply for vortioxetine. This needs to be considered when thinking about the potential generalization of these results to other populations.

In our simulations based on the real data, all Stage 1 methods (parametric and non-parametric) performed well according to prediction interval coverage of the true effects, across levels of heterogeneity in the effect across trials,  both a linear and non-linear CATE, a range of covariate distributions across the trials and target setting, and various numbers of studies. The performance of all methods varied across levels of age; specifically, coverage, length, and absolute bias were worse for many of the approaches and data generation setups for values of age that were further from the mean age. Therefore, in practice, it is important to explore the variability in the covariate distributions in the target setting and in comparison to the trials to identify if there might be subgroups for whom the CATE prediction intervals might perform more poorly. Furthermore, in the simulation setup where CATE coefficients across studies were pulled from a uniform distribution as opposed to a normal distribution, there remained high coverage but hovering below 95\%, indicating reason for some robustness but also caution regarding when the assumption of normality across studies is not met. 

In comparing Stage 1 methods, while there were not notable performance differences across methods on the whole, it is important to consider the different properties of each technique. Linear models stand separately to the other two approaches implemented here in that they are parametric and require pre-specification of hypothesized relationships and effect moderation. Causal forest and BART do not require this pre-specification but also do not have the same asymptotic properties as linear models. In comparing the two, the causal forest is a special case of the R-learner \citep{nie_quasi-oracle_2021} which has asymptotic efficiency benefits compared to the S-learner that was implemented with BART in this paper. The S-learner with BART has been shown to at times be biased towards zero \citep{kunzel2019metalearners} and have finite sample bias \citep{chernozhukov2018double}. While there were not substantial differences in the present simulations, these factors could affect prediction interval coverage in practice.


Notably, throughout this paper we have emphasized the idea of the CATE for a set of covariate profiles. This emphasis is first due to the nature of non-parametric, machine learning methods. Specifically, while meta-analysis provides an interpretable functional form of the CATE, approaches like the causal forest and BART instead give predicted CATEs and intervals but not a parametric form with coefficient estimates. Therefore, we represent the CATE by estimating it across a set of covariate profiles. Another key point is the important distinction between conditional average treatment effects (CATEs) and individual treatment effects (ITEs) \citep{post2025beyond, mueller2023personalized, post2024flexible}. The two estimands are often treated as the same but are not, and it is important to draw distinctions between them. An ITE is the difference between the potential outcomes for an individual and is very challenging (almost impossible) to estimate given that for each individual we only directly observe one of their potential outcomes. In contrast, the CATE represents average effects across individuals in the population that share the same observed covariates \citep{vegetabile2021distinction}, and it is more easily estimable. It is thus important to emphasize that the effects discussed in this paper are averages for a covariate profile rather than effect estimates for particular individuals.

The approaches discussed in this paper do have some limitations and opportunity for future refinement. Specifically, we rely on several assumptions to implement the methods, including an assumption of overlapping covariate distributions across the target and trial data (Assumption \ref{asx5}). A key concern might be predicting effects for covariate profiles in the health care system data that are not represented in the trials, which would violate our assumption. In our simulations, we set our target setting distributions to be different than those from our trials to assess this assumption. We saw that for the most part coverage was not an issue across the target setting, but that for individuals with more extreme ages (especially older ages that may have not been represented in the trials), coverage decreased. In the real data analysis, we removed covariate profiles from our target setting who had less than moderate depression, but there might be interest in understanding this group further. Future development could work to address this covariate overlap assumption; one approach could be augmenting the trial data with observational data that did have observed treatment and outcomes for this underrepresented group. Another important assumption of our approach is that the CATEs for a particular covariate profile $\boldsymbol{X}^*$ follow a normal distribution across studies/settings (Assumption \ref{asxnormal}). While this leverages the assumption employed for standard meta-analysis, it is important that researchers consider whether this assumption is plausible or if they may expect a non-normal distribution across studies. In addition, there may be some robustness to this assumption; as shown in the Supplemental simulations, there was high coverage still under a non-normal distribution but slightly below 95\%. This assumption also implies that the variability in the CATE between two given trials within a superpopulation (as depicted in Figure \ref{fig:assx}) is similar to the variability in the CATE between a trial and the target setting. Such an assumption requires careful thought, and it is worth caution before applying the approaches in this paper to a scenario where the target setting CATE may be very different from the trial CATEs and therefore not reasonably part of the same superpopulation.

Use of EHR data to represent a target setting can also come with other challenges, including measurement error. More specifically, the measures in the EHR data might be subjective, highly missing, or different measures of similar constructs compared to the trial data. Future work could explore imputation approaches to deal with missing data in this context or ways to leverage open text data to fill in more information about patient care. Furthermore, the issue of differential measures that was encountered where the EHR data measured depression using a different scale than the trial data is not an uncommon one. This was addressed in the present analysis by creation of a binary variable from both scales to separate into moderate or severe depression; another approach could have been to map the scales to one another continuously, or create standardized measures in a different way. These discordant measures can also arise in combining the trials; the trials used here had the same outcome measures, but many other applications of this approach might involve different trial outcomes and predictors that would need to be harmonized. Notably, while this paper utilized EHR data to represent patient profiles in the target setting, other sources of data could be used, which could introduce similar or new challenges as those that come with EHR data.

Future work should examine in more detail the choice of critical value in the prediction interval construction. We chose the $t$-distribution with $K-2$ degrees of freedom based on suggestions in the literature \citep{riley_interpretation_2011}. The true distribution of the treatment effects might not be approximately normal in reality, and future work could further refine the distributional assumptions and work to achieve close to 95\% coverage across non-parametric and parametric approaches. Another potential approach could be to explore Bayesian meta-analysis \citep{higgins_re-evaluation_2009} to leverage prior information about treatment effects when integrating information across trials. In general, precision in the prediction intervals can be increased with more studies and potentially incorporating strong non-experimental data. With the use of the $t$-statistic with $K-2$ degrees of freedom, more studies (higher values of $K$) will decrease the $t$-value and yield more narrow precision intervals.

Estimating heterogeneous treatment effects is of high interest in precision mental health and medicine as a whole, but doing so can be challenging statistically. Furthermore, practitioners are often interested in understanding the predicted effects before an individual has received any sort of treatment so that the treatment allocation can be optimal for patient outcomes. This paper introduced an approach for predicting effects in a target setting based on previously conducted trials of major depression medications. With further refinement of variance estimation and approaches for effectively dealing with measurement error and non-overlap in the EHR data, we can move closer towards aiding clinical decision-making based on which treatment is predicted to lead to better outcomes for the patient.

\section{Acknowledgments}

We would like to acknowledge Leon Di Stefano for his help sharing relevant references and giving feedback on the ideas and methods developed in this paper. The first author on this manuscript, Dr. Carly Lupton Brantner, completed some of the work on this paper as a PhD student at the Johns Hopkins Bloomberg School of Health.

This study was funded by the Patient-Centered Outcomes Research Institute (PCORI) through PCORI Award ME-2020C3-21145 (PI: Stuart) and the National Institute of Mental Health (NIMH) through Award R01MH126856 (PI: Stuart). Opinions and information in this content are those of the study authors and do not necessarily represent the views of PCORI or NIMH. Accordingly, PCORI and NIMH cannot make any guarantees with respect to the accuracy or reliability of the information and data.

Furthermore, this paper is based on research using data from data contributors, Takeda and Lundbeck, that has been made available through Vivli, Inc. Vivli has not contributed to or approved, and is not in any way responsible for, the contents of this publication. This study, carried out under YODA Project 2022-4854, used data obtained from the Yale University Open Data Access Project, which has an agreement with Janssen Research \& Development, L.L.C. The interpretation and reporting of research using this data are solely the responsibility of the authors and does not necessarily represent the official views of the Yale University Open Data Access Project or Janssen Research \& Development, L.L.C.

\textit{Conflict of interests:} None declared.

\section{Data Availability}

The data used in this paper, including data from \cite{mahableshwarkar_randomized_2013}, \cite{mahableshwarkar_randomized_2015}, \cite{boulenger_efficacy_2014}, and \cite{baldwin2012randomised} are only available through permission from the third party contributers and were accessed in this study via the Vivli platform. Simulated datasets can be replicated using code at the repository: \verb|https://github.com/carlyls/Predict_CATE|.

\begin{appendix}

\section{Bayesian Additive Regression Trees: CATE Estimation} \label{appendix_bart}

Bayesian additive regression trees (BART) is a sum-of-trees modeling procedure in conjunction with a regularization prior, where the prior restricts the amount that each tree can contribute to the overall model fit. An important distinction with BART is that it estimates the expected outcome conditional on covariates. To apply BART to estimate the CATE, several approaches exist, including the S-learner, T-learner \citep{kunzel2019metalearners}, and the Bayesian causal forest \citep{hahn_bayesian_2020}. We focus on the S-learner in this paper, as it is relatively straightforward to implement and applies well to the setting with multiple trials. We utilize the dbarts implementation \citep{dorie2023package} in this paper.

When using BART as an S-learner and in the setting with multiple trials, we fit a single model to each trial estimate $\mathbb{E}(Y|\boldsymbol{X}, A, S=s)$ -- the expected outcome given covariates and treatment in study $s$. As ``testing'' data, we replicate the training data but assign the opposite treatment to what the individual actually received, to estimate their counterfactual outcome. BART ultimately provides draws from the posterior distribution for the training outcomes under their true and counterfactual treatment assignment, and we can use these draws to estimate the treatment effects conditional on covariates and create credible intervals for these effect estimates. There are two options for how to estimate these credible intervals:

\begin{enumerate}
    \item We can estimate $\mathbb{E}(Y|\boldsymbol{X}, 1, S=s)$ and $\mathbb{E}(Y|\boldsymbol{X}, 0, S=s)$ by taking the average of the posterior draws for each treatment condition, and we can estimate $\text{Var}(Y|\boldsymbol{X}, 1, S=s)$ and $\text{Var}(Y|\boldsymbol{X}, 0, S=s)$ by taking the variance of the posterior draws for each treatment condition. We can then estimate the CATE by defining $\tau_s(\boldsymbol{X}) = \mathbb{E}(Y|\boldsymbol{X}, 1, S=s) - \mathbb{E}(Y|\boldsymbol{X}, 0, S=s)$, and we can estimate the variance of that CATE estimate by adding $\text{Var}(Y|\boldsymbol{X}, 1, S=s) + \text{Var}(Y|\boldsymbol{X}, 0, S=s)$, under the conservative assumption that the potential outcomes under treatment and control are uncorrelated. We  then create an interval under the assumption that $$\hat{\tau}_s(\boldsymbol{X}) \sim \mathcal{N}(\tau_s(\boldsymbol{X}), \text{Var}(Y|\boldsymbol{X}, 1, S=s) + \text{Var}(Y|\boldsymbol{X}, 0, S=s)).$$ 
    \item Instead of aggregating across posterior draws first, we can start by subtracting  $(\hat{Y}|\boldsymbol{X}, 1, S=s) - (\hat{Y}|\boldsymbol{X}, 0, S=s)$ within each posterior draw. We can then take the mean of these treatment effect estimates as well as the 2.5th and 97.5th percentiles of the CATE across these draws to construct a credible interval.
\end{enumerate}

We use option 1 above in this paper, but did explore both options in preliminary simulations. In our setting, we found that the more conservative option 1 performed better, especially when focusing on interval coverage in the target setting. Notably, the above options refer to constructing intervals in the original trials. In the target setting, we rely on estimating the treatment effect for the given covariate profile within each study, and then accounting for the within- and between-study variance to produce prediction intervals for the new setting (as discussed in Section \ref{stage2}).
\end{appendix}

\bibliographystyle{abbrvnat}
\bibliography{references}

\begin{thebibliography}{48}
\providecommand{\natexlab}[1]{#1}
\providecommand{\url}[1]{\texttt{#1}}
\expandafter\ifx\csname urlstyle\endcsname\relax
  \providecommand{\doi}[1]{doi: #1}\else
  \providecommand{\doi}{doi: \begingroup \urlstyle{rm}\Url}\fi

\bibitem[Athey et~al.(2019)Athey, Tibshirani, and Wager]{athey2019generalized}
S.~Athey, J.~Tibshirani, and S.~Wager.
\newblock Generalized random forests.
\newblock \emph{The Annals of Statistics}, 47\penalty0 (2):\penalty0 1148--1178, 2019.

\bibitem[Baldwin et~al.(2012)Baldwin, Loft, and Dragheim]{baldwin2012randomised}
D.~S. Baldwin, H.~Loft, and M.~Dragheim.
\newblock A randomised, double-blind, placebo controlled, duloxetine-referenced, fixed-dose study of three dosages of lu aa21004 in acute treatment of major depressive disorder (mdd).
\newblock \emph{European Neuropsychopharmacology}, 22\penalty0 (7):\penalty0 482--491, 2012.

\bibitem[Boulenger et~al.(2014)Boulenger, Loft, and Olsen]{boulenger_efficacy_2014}
J.-P. Boulenger, H.~Loft, and C.~K. Olsen.
\newblock Efficacy and safety of vortioxetine ({Lu} {AA21004}), 15 and 20 mg/day: a randomized, double-blind, placebo-controlled, duloxetine-referenced study in the acute treatment of adult patients with major depressive disorder.
\newblock \emph{International Clinical Psychopharmacology}, 29\penalty0 (3):\penalty0 138--149, May 2014.
\newblock ISSN 0268-1315.
\newblock \doi{10.1097/YIC.0000000000000018}.
\newblock URL \url{http://journals.lww.com/00004850-201405000-00002}.

\bibitem[Brantner et~al.(2023)Brantner, Chang, Nguyen, Hong, Di~Stefano, and Stuart]{brantner2023review}
C.~L. Brantner, T.-H. Chang, T.~Q. Nguyen, H.~Hong, L.~Di~Stefano, and E.~A. Stuart.
\newblock Methods for integrating trials and non-experimental data to examine treatment effect heterogeneity.
\newblock \emph{Statistical Science}, 38\penalty0 (4):\penalty0 640--654, 2023.

\bibitem[Brantner et~al.(2024)Brantner, Nguyen, Tang, Zhao, Hong, and Stuart]{brantner2024comparison}
C.~L. Brantner, T.~Q. Nguyen, T.~Tang, C.~Zhao, H.~Hong, and E.~A. Stuart.
\newblock Comparison of methods that combine multiple randomized trials to estimate heterogeneous treatment effects.
\newblock \emph{Statistics in Medicine}, 2024.

\bibitem[Burke et~al.(2017)Burke, Ensor, and Riley]{burke_meta-analysis_2017}
D.~L. Burke, J.~Ensor, and R.~D. Riley.
\newblock Meta-analysis using individual participant data: one-stage and two-stage approaches, and why they may differ.
\newblock \emph{Statistics in Medicine}, 36\penalty0 (5):\penalty0 855--875, feb 2017.
\newblock ISSN 02776715.
\newblock \doi{10.1002/sim.7141}.
\newblock URL \url{https://onlinelibrary.wiley.com/doi/10.1002/sim.7141}.

\bibitem[Carnegie et~al.(2019)Carnegie, Dorie, and Hill]{carnegie_examining_2019}
N.~Carnegie, V.~Dorie, and J.~L. Hill.
\newblock Examining treatment effect heterogeneity using {BART}.
\newblock \emph{Observational Studies}, 5\penalty0 (2):\penalty0 52--70, 2019.
\newblock ISSN 2767-3324.
\newblock \doi{10.1353/obs.2019.0002}.
\newblock URL \url{https://muse.jhu.edu/article/793357}.

\bibitem[Chernozhukov et~al.(2018)Chernozhukov, Chetverikov, Demirer, Duflo, Hansen, Newey, and Robins]{chernozhukov2018double}
V.~Chernozhukov, D.~Chetverikov, M.~Demirer, E.~Duflo, C.~Hansen, W.~Newey, and J.~Robins.
\newblock Double/debiased machine learning for treatment and structural parameters, 2018.

\bibitem[Colnet et~al.(2021)Colnet, Mayer, Chen, Dieng, Li, Varoquaux, Vert, Josse, and Yang]{colnet_causal_2021}
B.~Colnet, I.~Mayer, G.~Chen, A.~Dieng, R.~Li, G.~Varoquaux, J.-P. Vert, J.~Josse, and S.~Yang.
\newblock Causal inference methods for combining randomized trials and observational studies: a review.
\newblock \emph{arXiv:2011.08047 [stat]}, 2021.
\newblock URL \url{http://arxiv.org/abs/2011.08047}.
\newblock arXiv: 2011.08047.

\bibitem[Currie and MacLeod(2020)]{currie2020understanding}
J.~M. Currie and W.~B. MacLeod.
\newblock Understanding doctor decision making: The case of depression treatment.
\newblock \emph{Econometrica}, 88\penalty0 (3):\penalty0 847--878, 2020.

\bibitem[Dahabreh et~al.(2020)Dahabreh, Petito, Robertson, Hernán, and Steingrimsson]{dahabreh_towards_2020}
I.~J. Dahabreh, L.~C. Petito, S.~E. Robertson, M.~A. Hernán, and J.~A. Steingrimsson.
\newblock Towards causally interpretable meta-analysis: transporting inferences from multiple studies to a target population.
\newblock \emph{arXiv:1903.11455 [stat]}, 2020.
\newblock URL \url{http://arxiv.org/abs/1903.11455}.
\newblock arXiv: 1903.11455.

\bibitem[DerSimonian and Laird(1986)]{dersimonian1986meta}
R.~DerSimonian and N.~Laird.
\newblock Meta-analysis in clinical trials.
\newblock \emph{Controlled clinical trials}, 7\penalty0 (3):\penalty0 177--188, 1986.

\bibitem[Dhaliwal et~al.(2023)Dhaliwal, Spurling, and Molla]{Dhaliwal_Spurling_Molla_2023}
J.~S.~S. Dhaliwal, B.~C. Spurling, and M.~Molla.
\newblock Duloxetine, May 2023.
\newblock URL \url{https://www.ncbi.nlm.nih.gov/books/NBK549806/}.

\bibitem[Dorie et~al.(2022)Dorie, Perrett, Hill, and Goodrich]{dorie_stan_2022}
V.~Dorie, G.~Perrett, J.~L. Hill, and B.~Goodrich.
\newblock Stan and {BART} for {Causal} {Inference}: {Estimating} {Heterogeneous} {Treatment} {Effects} {Using} the {Power} of {Stan} and the {Flexibility} of {Machine} {Learning}.
\newblock \emph{Entropy}, 24\penalty0 (12):\penalty0 1782, Dec. 2022.
\newblock ISSN 1099-4300.
\newblock \doi{10.3390/e24121782}.
\newblock URL \url{https://www.ncbi.nlm.nih.gov/pmc/articles/PMC9778579/}.

\bibitem[Dorie et~al.(2023)Dorie, Chipman, McCulloch, Dadgar, Team, Draheim, Bosmans, Tournayre, Petch, de~Lucena~Valle, et~al.]{dorie2023package}
V.~Dorie, H.~Chipman, R.~McCulloch, A.~Dadgar, R.~C. Team, G.~U. Draheim, M.~Bosmans, C.~Tournayre, M.~Petch, R.~de~Lucena~Valle, et~al.
\newblock Package ‘dbarts’.
\newblock 2023.

\bibitem[Dzau and Ginsburg(2016)]{dzau2016realizing}
V.~J. Dzau and G.~S. Ginsburg.
\newblock Realizing the full potential of precision medicine in health and health care.
\newblock \emph{Jama}, 316\penalty0 (16):\penalty0 1659--1660, 2016.

\bibitem[D’Agostino et~al.(2015)D’Agostino, English, and Rey]{d2015vortioxetine}
A.~D’Agostino, C.~D. English, and J.~A. Rey.
\newblock Vortioxetine (brintellix): a new serotonergic antidepressant.
\newblock \emph{Pharmacy and Therapeutics}, 40\penalty0 (1):\penalty0 36, 2015.

\bibitem[Girardi et~al.(2009)Girardi, Pompili, Innamorati, Mancini, Serafini, Mazzarini, Del~Casale, Tatarelli, and Baldessarini]{girardi2009duloxetine}
P.~Girardi, M.~Pompili, M.~Innamorati, M.~Mancini, G.~Serafini, L.~Mazzarini, A.~Del~Casale, R.~Tatarelli, and R.~J. Baldessarini.
\newblock Duloxetine in acute major depression: review of comparisons to placebo and standard antidepressants using dissimilar methods.
\newblock \emph{Human Psychopharmacology: Clinical and Experimental}, 24\penalty0 (3):\penalty0 177--190, 2009.

\bibitem[Goldstein et~al.(2004)Goldstein, Lu, Detke, Wiltse, Mallinckrodt, and Demitrack]{goldstein2004duloxetine}
D.~J. Goldstein, Y.~Lu, M.~J. Detke, C.~Wiltse, C.~Mallinckrodt, and M.~A. Demitrack.
\newblock Duloxetine in the treatment of depression: a double-blind placebo-controlled comparison with paroxetine.
\newblock \emph{Journal of clinical psychopharmacology}, 24\penalty0 (4):\penalty0 389--399, 2004.

\bibitem[Hahn et~al.(2020)Hahn, Murray, and Carvalho]{hahn_bayesian_2020}
P.~R. Hahn, J.~S. Murray, and C.~M. Carvalho.
\newblock Bayesian {Regression} {Tree} {Models} for {Causal} {Inference}: {Regularization}, {Confounding}, and {Heterogeneous} {Effects} (with {Discussion}).
\newblock \emph{Bayesian Analysis}, 15\penalty0 (3):\penalty0 965--1056, Sept. 2020.
\newblock ISSN 1936-0975, 1931-6690.
\newblock \doi{10.1214/19-BA1195}.
\newblock URL \url{https://projecteuclid.org/journals/bayesian-analysis/volume-15/issue-3/Bayesian-Regression-Tree-Models-for-Causal-Inference--Regularization-Confounding/10.1214/19-BA1195.full}.
\newblock Publisher: International Society for Bayesian Analysis.

\bibitem[Herrmann et~al.(1998)Herrmann, Black, Lawrence, Szekely, and Szalai]{herrmann1998sunnybrook}
N.~Herrmann, S.~Black, J.~Lawrence, C.~Szekely, and J.~Szalai.
\newblock The sunnybrook stroke study: a prospective study of depressive symptoms and functional outcome.
\newblock \emph{Stroke}, 29\penalty0 (3):\penalty0 618--624, 1998.

\bibitem[Higgins et~al.(2009)Higgins, Thompson, and Spiegelhalter]{higgins_re-evaluation_2009}
J.~P.~T. Higgins, S.~G. Thompson, and D.~J. Spiegelhalter.
\newblock A re-evaluation of random-effects meta-analysis.
\newblock \emph{Journal of the Royal Statistical Society: Series A (Statistics in Society)}, 172\penalty0 (1):\penalty0 137--159, 2009.
\newblock ISSN 1467-985X.
\newblock \doi{10.1111/j.1467-985X.2008.00552.x}.
\newblock URL \url{https://onlinelibrary.wiley.com/doi/abs/10.1111/j.1467-985X.2008.00552.x}.
\newblock \_eprint: https://onlinelibrary.wiley.com/doi/pdf/10.1111/j.1467-985X.2008.00552.x.

\bibitem[Hill(2011)]{hill_bayesian_2011}
J.~L. Hill.
\newblock Bayesian {Nonparametric} {Modeling} for {Causal} {Inference}.
\newblock \emph{Journal of Computational and Graphical Statistics}, 20\penalty0 (1):\penalty0 217--240, Jan. 2011.
\newblock ISSN 1061-8600, 1537-2715.
\newblock \doi{10.1198/jcgs.2010.08162}.
\newblock URL \url{http://www.tandfonline.com/doi/abs/10.1198/jcgs.2010.08162}.

\bibitem[IntHout et~al.(2016)IntHout, Ioannidis, Rovers, and Goeman]{inthout2016plea}
J.~IntHout, J.~P. Ioannidis, M.~M. Rovers, and J.~J. Goeman.
\newblock Plea for routinely presenting prediction intervals in meta-analysis.
\newblock \emph{BMJ open}, 6\penalty0 (7):\penalty0 e010247, 2016.

\bibitem[Kosorok and Laber(2019)]{kosorok2019precision}
M.~R. Kosorok and E.~B. Laber.
\newblock Precision medicine.
\newblock \emph{Annual review of statistics and its application}, 6\penalty0 (1):\penalty0 263--286, 2019.

\bibitem[Kroenke et~al.(2001)Kroenke, Spitzer, and Williams]{kroenke2001phq}
K.~Kroenke, R.~L. Spitzer, and J.~B. Williams.
\newblock The phq-9: validity of a brief depression severity measure.
\newblock \emph{Journal of general internal medicine}, 16\penalty0 (9):\penalty0 606--613, 2001.

\bibitem[K{\"u}nzel et~al.(2019)K{\"u}nzel, Sekhon, Bickel, and Yu]{kunzel2019metalearners}
S.~R. K{\"u}nzel, J.~S. Sekhon, P.~J. Bickel, and B.~Yu.
\newblock Metalearners for estimating heterogeneous treatment effects using machine learning.
\newblock \emph{Proceedings of the national academy of sciences}, 116\penalty0 (10):\penalty0 4156--4165, 2019.

\bibitem[Li et~al.(2016)Li, Wang, and Ma]{li2016vortioxetine}
G.~Li, X.~Wang, and D.~Ma.
\newblock Vortioxetine versus duloxetine in the treatment of patients with major depressive disorder: a meta-analysis of randomized controlled trials.
\newblock \emph{Clinical drug investigation}, 36:\penalty0 509--517, 2016.

\bibitem[Mahableshwarkar et~al.(2013)Mahableshwarkar, Jacobsen, and Chen]{mahableshwarkar_randomized_2013}
A.~R. Mahableshwarkar, P.~L. Jacobsen, and Y.~Chen.
\newblock A randomized, double-blind trial of 2.5 mg and 5 mg vortioxetine ({Lu} {AA21004}) versus placebo for 8 weeks in adults with major depressive disorder.
\newblock \emph{Current Medical Research and Opinion}, 29\penalty0 (3):\penalty0 217--226, Mar. 2013.
\newblock ISSN 0300-7995, 1473-4877.
\newblock \doi{10.1185/03007995.2012.761600}.
\newblock URL \url{http://www.tandfonline.com/doi/full/10.1185/03007995.2012.761600}.

\bibitem[Mahableshwarkar et~al.(2015)Mahableshwarkar, Jacobsen, Chen, Serenko, and Trivedi]{mahableshwarkar_randomized_2015}
A.~R. Mahableshwarkar, P.~L. Jacobsen, Y.~Chen, M.~Serenko, and M.~H. Trivedi.
\newblock A randomized, double-blind, duloxetine-referenced study comparing efficacy and tolerability of 2 fixed doses of vortioxetine in the acute treatment of adults with {MDD}.
\newblock \emph{Psychopharmacology}, 232\penalty0 (12):\penalty0 2061--2070, June 2015.
\newblock ISSN 0033-3158, 1432-2072.
\newblock \doi{10.1007/s00213-014-3839-0}.
\newblock URL \url{http://link.springer.com/10.1007/s00213-014-3839-0}.

\bibitem[Mills et~al.(2021)Mills, Higgins, Morris, Kessler, Heron, Wiles, Smith, and Tilling]{mills2021detecting}
H.~L. Mills, J.~P. Higgins, R.~W. Morris, D.~Kessler, J.~Heron, N.~Wiles, G.~D. Smith, and K.~Tilling.
\newblock Detecting heterogeneity of intervention effects using analysis and meta-analysis of differences in variance between trial arms.
\newblock \emph{Epidemiology (Cambridge, Mass.)}, 32\penalty0 (6):\penalty0 846, 2021.

\bibitem[Montgomery and {\AA}sberg(1979)]{montgomery1979new}
S.~A. Montgomery and M.~{\AA}sberg.
\newblock A new depression scale designed to be sensitive to change.
\newblock \emph{The British journal of psychiatry}, 134\penalty0 (4):\penalty0 382--389, 1979.

\bibitem[Mueller and Pearl(2023)]{mueller2023personalized}
S.~Mueller and J.~Pearl.
\newblock Personalized decision making--a conceptual introduction.
\newblock \emph{Journal of Causal Inference}, 11\penalty0 (1):\penalty0 20220050, 2023.

\bibitem[Nie and Wager(2021)]{nie_quasi-oracle_2021}
X.~Nie and S.~Wager.
\newblock Quasi-oracle estimation of heterogeneous treatment effects.
\newblock \emph{Biometrika}, 108\penalty0 (2):\penalty0 299--319, 2021.
\newblock ISSN 0006-3444, 1464-3510.
\newblock \doi{10.1093/biomet/asaa076}.
\newblock URL \url{https://academic.oup.com/biomet/article/108/2/299/5911092}.

\bibitem[NIMH(2024)]{NIMH_2024}
NIMH.
\newblock The national institute of mental health strategic plan, May 2024.
\newblock URL \url{https://www.nimh.nih.gov/about/strategic-planning-reports}.

\bibitem[Post and Van Den~Heuvel(2025)]{post2025beyond}
R.~A. Post and E.~R. Van Den~Heuvel.
\newblock Beyond conditional averages: Estimating the individual causal effect distribution.
\newblock \emph{Journal of Causal Inference}, 13\penalty0 (1):\penalty0 20240007, 2025.

\bibitem[Post et~al.(2024)Post, Petkovic, Van~den Heuvel, and Van~den Heuvel]{post2024flexible}
R.~A. Post, M.~Petkovic, I.~L. Van~den Heuvel, and E.~R. Van~den Heuvel.
\newblock Flexible machine learning estimation of conditional average treatment effects: a blessing and a curse.
\newblock \emph{Epidemiology}, 35\penalty0 (1):\penalty0 32--40, 2024.

\bibitem[Riley et~al.(2011)Riley, Higgins, and Deeks]{riley_interpretation_2011}
R.~D. Riley, J.~P.~T. Higgins, and J.~J. Deeks.
\newblock Interpretation of random effects meta-analyses.
\newblock \emph{BMJ}, 342\penalty0 (feb10 2):\penalty0 d549--d549, Feb. 2011.
\newblock ISSN 0959-8138, 1468-5833.
\newblock \doi{10.1136/bmj.d549}.
\newblock URL \url{https://www.bmj.com/lookup/doi/10.1136/bmj.d549}.

\bibitem[Riley et~al.(2021)Riley, Debray, Morris, and Jackson]{riley_two-stage_2021}
R.~D. Riley, T.~P. Debray, T.~P. Morris, and D.~Jackson.
\newblock The {Two}-stage {Approach} to {IPD} {Meta}-{Analysis}.
\newblock In \emph{Individual {Participant} {Data} {Meta}-{Analysis}}, pages 87--125. John Wiley \& Sons, Ltd, 2021.
\newblock ISBN 978-1-119-33378-4.
\newblock \doi{10.1002/9781119333784.ch5}.
\newblock URL \url{https://onlinelibrary.wiley.com/doi/abs/10.1002/9781119333784.ch5}.
\newblock Section: 5 \_eprint: https://onlinelibrary.wiley.com/doi/pdf/10.1002/9781119333784.ch5.

\bibitem[Roth and Fonagy(2006)]{roth2006works}
A.~Roth and P.~Fonagy.
\newblock What works for whom?: a critical review of psychotherapy research.
\newblock 2006.

\bibitem[Rubin(1974)]{rubin_estimating_1974}
D.~B. Rubin.
\newblock Estimating causal effects of treatments in randomized and nonrandomized studies.
\newblock \emph{Journal of Educational Psychology}, 66\penalty0 (5):\penalty0 688--701, 1974.
\newblock ISSN 1939-2176.
\newblock \doi{10.1037/h0037350}.
\newblock Place: US Publisher: American Psychological Association.

\bibitem[Sheffler et~al.(2023)Sheffler, Patel, and Abdijadid]{Sheffler_Patel_Abdijadid_2023}
Z.~M. Sheffler, P.~Patel, and S.~Abdijadid.
\newblock Antidepressants, May 2023.
\newblock URL \url{https://www.ncbi.nlm.nih.gov/books/NBK538182/}.

\bibitem[Snaith et~al.(1986)Snaith, Harrop, Newby, and Teale]{snaith1986grade}
R.~Snaith, F.~Harrop, t.~D. Newby, and C.~Teale.
\newblock Grade scores of the montgomery—{\aa}sberg depression and the clinical anxiety scales.
\newblock \emph{The British journal of psychiatry}, 148\penalty0 (5):\penalty0 599--601, 1986.

\bibitem[Sobel et~al.(2017)Sobel, Madigan, and Wang]{sobel_causal_2017}
M.~Sobel, D.~Madigan, and W.~Wang.
\newblock Causal {Inference} for {Meta}-{Analysis} and {Multi}-{Level} {Data} {Structures}, with {Application} to {Randomized} {Studies} of {Vioxx}.
\newblock \emph{Psychometrika}, 82\penalty0 (2):\penalty0 459--474, 2017.
\newblock ISSN 0033-3123, 1860-0980.
\newblock \doi{10.1007/s11336-016-9507-z}.
\newblock URL \url{http://link.springer.com/10.1007/s11336-016-9507-z}.

\bibitem[Vegetabile(2021)]{vegetabile2021distinction}
B.~G. Vegetabile.
\newblock On the distinction between" conditional average treatment effects"(cate) and" individual treatment effects"(ite) under ignorability assumptions.
\newblock \emph{arXiv preprint arXiv:2108.04939}, 2021.

\bibitem[Viechtbauer(2010)]{metafor}
W.~Viechtbauer.
\newblock Conducting meta-analyses in {R} with the {metafor} package.
\newblock \emph{Journal of Statistical Software}, 36\penalty0 (3):\penalty0 1--48, 2010.
\newblock \doi{10.18637/jss.v036.i03}.

\bibitem[Wager and Athey(2018)]{wager2018estimation}
S.~Wager and S.~Athey.
\newblock Estimation and inference of heterogeneous treatment effects using random forests.
\newblock \emph{Journal of the American Statistical Association}, 113\penalty0 (523):\penalty0 1228--1242, 2018.

\bibitem[Yusuf et~al.(1991)Yusuf, Wittes, Probstfield, and Tyroler]{yusuf1991analysis}
S.~Yusuf, J.~Wittes, J.~Probstfield, and H.~A. Tyroler.
\newblock Analysis and interpretation of treatment effects in subgroups of patients in randomized clinical trials.
\newblock \emph{Jama}, 266\penalty0 (1):\penalty0 93--98, 1991.

\end{thebibliography}

\end{document}